\documentclass{aastex631}
\pdfoutput=1

\usepackage{amssymb,amsmath}
\usepackage{soul}

\usepackage{needspace}
\usepackage{listings}

\definecolor{codegreen}{rgb}{0,0.6,0}
\definecolor{codegray}{rgb}{0.5,0.5,0.5}
\definecolor{codepurple}{rgb}{0.58,0,0.82}
\definecolor{backcolour}{rgb}{0.98,0.98,0.95}

\lstdefinestyle{mystyle}{
    backgroundcolor=\color{backcolour},   
    commentstyle=\color{codegreen},
    keywordstyle=\color{magenta},
    numberstyle=\tiny\color{codegray},
    stringstyle=\color{codepurple},
    basicstyle=\ttfamily\footnotesize,
    breakatwhitespace=false,         
    breaklines=true,                 
    captionpos=b,                    
    keepspaces=true,                 
    numbers=left,                    
    numbersep=5pt,                  
    showspaces=false,                
    showstringspaces=false,
    showtabs=false,                  
    tabsize=2
}

\defcitealias{Brandt_2024}{Paper II}

\begin{document}

\title{Optimal Fitting and Debiasing for Detectors Read Out Up-the-Ramp}

\author[0000-0003-2630-8073]{Timothy D.~Brandt}
\affiliation{Space Telescope Science Institute \\ 3700 San Martin Drive \\ Baltimore, MD 21218, USA}
\affiliation{Department of Physics, University of California, Santa Barbara \\ Broida Hall \\ Santa Barbara, CA, 93106, USA}

\begin{abstract}

This paper derives the optimal fit to a pixel's count rate in the case of an ideal detector read out nondestructively in the presence of both read and photon noise.  The approach is general for any readout scheme, provides closed-form expressions for all quantities, and has a computational cost that is linear in the number of resultants (groups of reads).  I also derive the bias of the fit from estimating the covariance matrix and show how to remove it to first order.  The ramp-fitting algorithm I describe provides the $\chi^2$ value of the fit of a line to the accumulated counts, which can be interpreted as a goodness-of-fit metric.  I provide and describe a pure Python implementation of these algorithms that can process a 10-resultant ramp on a $4096 \times 4096$ detector in $\approx$8 seconds with bias removal on a single core of a 2020 Macbook Air.  This Python implementation, together with tests and a tutorial notebook, are available at \url{https://github.com/t-brandt/fitramp}.  A companion paper describes a jump detection algorithm based on hypothesis testing of ramp fits and demonstrates all algorithms on data from JWST.

\end{abstract}

\keywords{}

\section{Introduction and Statement of the problem} \label{sec:intro}

Many detectors may be read out nondestructively to reduce the impact of read noise, with the reads being saved either individually or in groups for later analysis.  This approach is standard on NICMOS \citep{Skinner+Bergeron+Schultz+etal_1998} and on the infrared channel of WFC3 \citep{Baggett+Hill+Kimble+etal_2008}, both of which are installed on the Hubble Space Telescope.  Ground-based instruments using infrared detectors can also be read out nondestructively.  Some save only a combination of the reads as an estimate of the count rate, while others save all individual reads.  The CHARIS instrument on the Subaru telescope is an example of the latter \citep{Groff+Chilcote+Kasdin+etal_2016,Brandt+Rizzo+Groff+etal_2017}.  

The initial phase of processing data from a detector read out nondestructively is to derive the count rate from a sequence of reads.  Each read measures the number of electrons in a pixel; it is subject to both read noise and photon noise.  For an ideal detector in the absence of read noise and photon noise, the number of counts in a pixel would be the reset value plus the count rate times the time since reset.  The reset value itself is subject to $kTC$ noise and must be fitted from the data.

The problem of fitting a ramp has been studied extensively in the past.  \cite{Fixsen+Offenberg+Hanisch+etal_2000} and \cite{Offenberg+Fixsen+Rauscher+etal_2001} derived and validated nearly optimal weights for combining individual, equally spaced reads as a function of signal-to-noise ratio.  They also used the individual, saved reads to identify cosmic rays as instantaneous jumps in a  pixel's counts.  \cite{Kubik+Barbier+Chabanat+etal_2016} extended the ramp fitting approach for the Euclid spacecraft while \cite{Casertano_2022} updated the weight calculation of \cite{Fixsen+Offenberg+Hanisch+etal_2000} for nonuniform sampling. 
\cite{Robberto_2014} proposed an optimal approach for ramp fitting at the cost of additional matrix operations to diagonalize each pixel's covariance matrix.  

In this work I revisit the problem of fitting a ramp to a sequence of nondestructive reads.  In the companion paper \cite{Brandt_2024}, hereafter \citetalias{Brandt_2024}, I address the problem of identifying jumps in a pixel's counts.  I consider the general case of a detector reading out at many arbitrary times and possibly averaging some of these reads together into groups; the average of a group of reads is also called a resultant.  The reads are typically averaged with equal weights.  Appendix \ref{sec:intraresultantweights} shows that equal weights are not optimal, and demonstrates the gains that are possible with alternative weights of the reads that combine to form a resultant.

Fitting a ramp to a sequence of resultants can be decomposed into two tasks.  The first task is to derive the covariance matrix for the resultants.  In practice, the read noise for each pixel may be measured, but the photon noise will have to be approximated from the data themselves.  The second task is to use the covariance matrix to derive the maximum likelihood count rate.  

The treatment I present here assumes an ideal detector and a constant astrophysical+dark count rate.  I further assume that shot noise, digitization noise, and other noise sources are sufficiently modeled as Gaussian rather than, e.g., Poisson.  This is necessary in order to identify the $\chi^2$ statistic with the log likelihood for hypothesis testing.  The assumption of an ideal detector includes perfect linearity and read noise that is uncorrelated between detector reads (though not necessarily between detector pixels).  The $1/f$ noise ubiquitous in H2RG infrared detectors \citep{Moseley+Arendt+Fixsen+etal_2010,Rauscher_2015} is not a problem so long as the very low frequency component ($\sim$seconds long, between reads at a fixed pixel) may be removed.  Deviations from linearity may be corrected to create a ramp appropriate for the treatment presented here.  Real detectors will have a number of additional complications, from pre-amplifier effects to random telegraph noise \citep[e.g.][]{Schlawin+Leisenring+Misselt+etal_2020}, that may or may not have a significant impact on the efficacy of the method presented here.

Consider a ramp consisting of many resultants ${\bf r}$.  If the covariance matrix ${\bf C}$ for this set of resultants is known, the problem of deriving the count rate involves minimizing
\begin{equation}
\chi^2 \equiv \left({\bf r}_{\rm meas} - {\bf r}_{\rm model} \right)^T {\bf C}^{-1} \left({\bf r}_{\rm meas} - {\bf r}_{\rm model} \right) \label{eq:chisq_general}
\end{equation}
where ${\bf r}_{\rm meas}$ are the measured counts in each resultant, ${\bf r}_{\rm model}$ are the model counts, and $\chi^2$ is the chi-squared statistic.  If a resultant consists of a single read, then
\begin{equation}
    r_{{\rm model}, i} = a t_i + b
\end{equation}
where $a$ is the count rate, $t_i$ is the time of that read (with $t=0$ corresponding to the time of the last reset), and $b$ is the reset value. 

Equation \eqref{eq:chisq_general} requires computing and then inverting a covariance matrix.  If the covariance matrix is dense, then its inversion lacks a convenient closed form and has a computational cost that scales as $n^3$, where $n$ is the number of resultants.  If this can be overcome, Section \ref{sec:improvementpotential} shows the potential improvement in signal-to-noise ratio over current, approximate approaches.

In the rest of this paper I will recast the problem using only the differences between resultants.  In Section \ref{sec:covmatrix} I will derive the resulting covariance matrix and show that it is tridiagonal.  In Section \ref{sec:fitramp} I will derive closed-form expressions for the maximum likelihood count rate and for the goodness-of-fit that may be used for hypothesis testing, e.g., for a possible jump in counts due to a cosmic ray hit.  In Section  \ref{sec:bias} I will derive an analytic expression for the first-order bias of the count rate estimator.  Section \ref{sec:implementation} describes a pure Python implementation of optimal ramp fitting at a cost linear in the number of resultants; it is computationally straightforward on a laptop computer even for long ramps on large-format detectors. I conclude with Section \ref{sec:conclusions}.

\section{Generalized Least Squares vs.~Approximate Approaches} \label{sec:improvementpotential}

The current data processing pipelines for HST and JWST use adaptations of the approach suggested by \cite{Fixsen+Offenberg+Hanisch+etal_2000} and \cite{Offenberg+Fixsen+Rauscher+etal_2001}.  This approach uses a weighted average of the resultants, where the weights are constant in bins of estimated signal-to-noise ratio.  The resulting weighted sum provides an estimate of the count rate that approaches, but does not reach, the precision of the treatment with the full covariance matrix.  Because the weights change discretely with the properties of a ramp, I refer to the approach of \cite{Fixsen+Offenberg+Hanisch+etal_2000} and \cite{Offenberg+Fixsen+Rauscher+etal_2001} as the discrete weighting case.  The full covariance matrix provides for continuously variable weights.

\begin{figure*}
\includegraphics[width=0.5\textwidth]{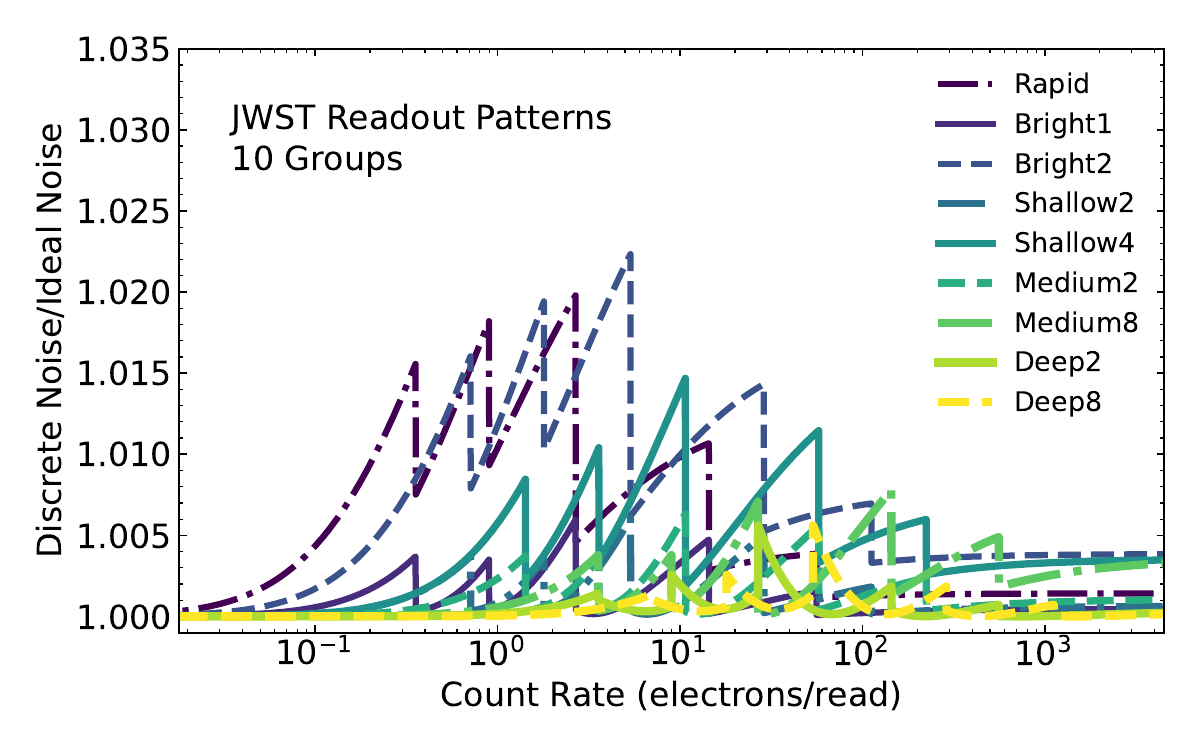}
\includegraphics[width=0.5\textwidth]{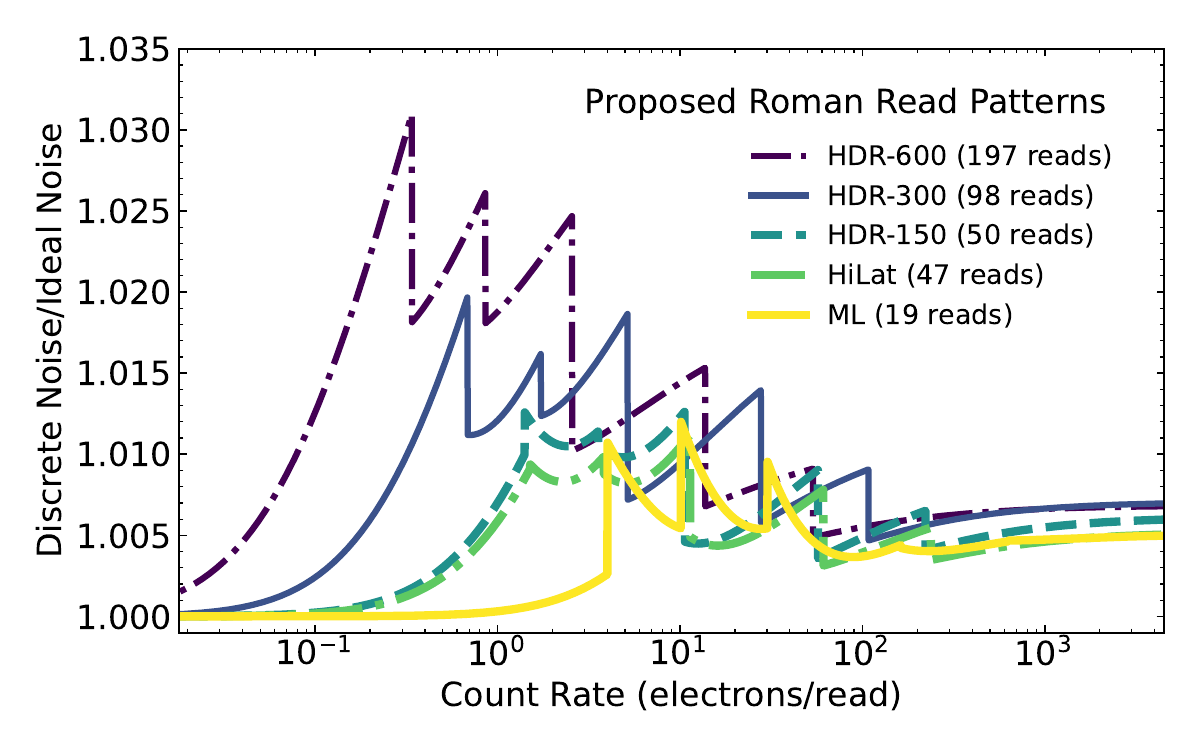}
\caption{Ratio of the noise in the count rate using the \cite{Offenberg+Fixsen+Rauscher+etal_2001} approach used for JWST (left) and the suggested modification by \cite{Casertano_2022} for Roman (right) to the noise from the $\chi^2$ approach using the full covariance matrix.  All current readout patterns of JWST are shown assuming 10 groups.  In both cases I assume a read noise of 10 electrons/read.  A fit using the full covariance matrix offers an improvement of up to $\approx$2\% in signal-to-noise for JWST and between 0.5\% and 3\% for Roman, corresponding to increases in collecting area of up to 4\% and between 1\% and 6\%, respectively.  The ``saw-tooth'' pattern comes from transitions between discrete weighting schemes that are close to, but not exactly at, the level where both weighting schemes produce the same signal-to-noise ratio on the ramp fit.
\label{fig:snr_compare}}
\end{figure*}

The most straightforward metric of the benefit of the full covariance matrix treatment presented in this paper is the signal-to-noise ratio of the inferred count rates.  Figure \ref{fig:snr_compare} shows the noise in the count rate for the approach of \cite{Fixsen+Offenberg+Hanisch+etal_2000} and \cite{Offenberg+Fixsen+Rauscher+etal_2001}, as adapted by \cite{Casertano_2022}, as a fraction of the noise from a $\chi^2$ minimization using the correct covariance matrix.  The latter approach provides the smallest uncertainty of all unbiased estimates; the ratios are strictly larger than one.  All currently available readout patterns for NIRCam on JWST\footnote{\url{https://jwst-docs.stsci.edu/jwst-near-infrared-camera/nircam-instrumentation/nircam-detector-overview/nircam-detector-readout-patterns}} are shown with 10 groups; the proposed Roman readout patterns are detailed in \cite{Casertano_2022}.  In all cases the covariance matrix itself is assumed to be known.  In reality the covariance matrix must be estimated; this introduces biases that I derive in Section \ref{sec:bias}.   I assume a fiducial read noise of 10 electrons in a single read ($10\sqrt{2}$ electrons in a read difference).  The noise in the discrete weighting case shows discontinuities where one set of weights transitions to another; this appears as a saw-tooth pattern on Figure \ref{fig:snr_compare}.  These discontinuities could be avoided by choosing slightly different signal-to-noise thresholds between the weighting schemes, such that both sets of weights produce the same signal-to-noise ratio at a transition.  Achieving this in practice would require the transitions between weights to change with the readout pattern.

The noise values shown in Figure \ref{fig:snr_compare} show improvements from the discrete weighting case ranging from $\ll$1\% for long JWST ramps with many reads per ramp (Deep8, with ten resultants each of eight reads) to 3\% for long Roman exposures from using a fit with the full covariance matrix.  Typical improvements range from 0.5\% to 2\%, corresponding to an increase in equivalent collecting area of 1\% to 4\%.  These represent meaningful improvements to the missions if the full $\chi^2$ fit can be implemented robustly and efficiently.  If the construction of the resultants themselves can be modified to accommodate weighted averages of reads, Appendix \ref{sec:intraresultantweights} shows that further improvements are possible especially with low numbers of resultants as would be the case if downlink bandwidth were severely restricted.

\section{Deriving the Covariance Matrix} \label{sec:covmatrix}

\begin{figure}
    \centering\includegraphics[width=0.5\textwidth]{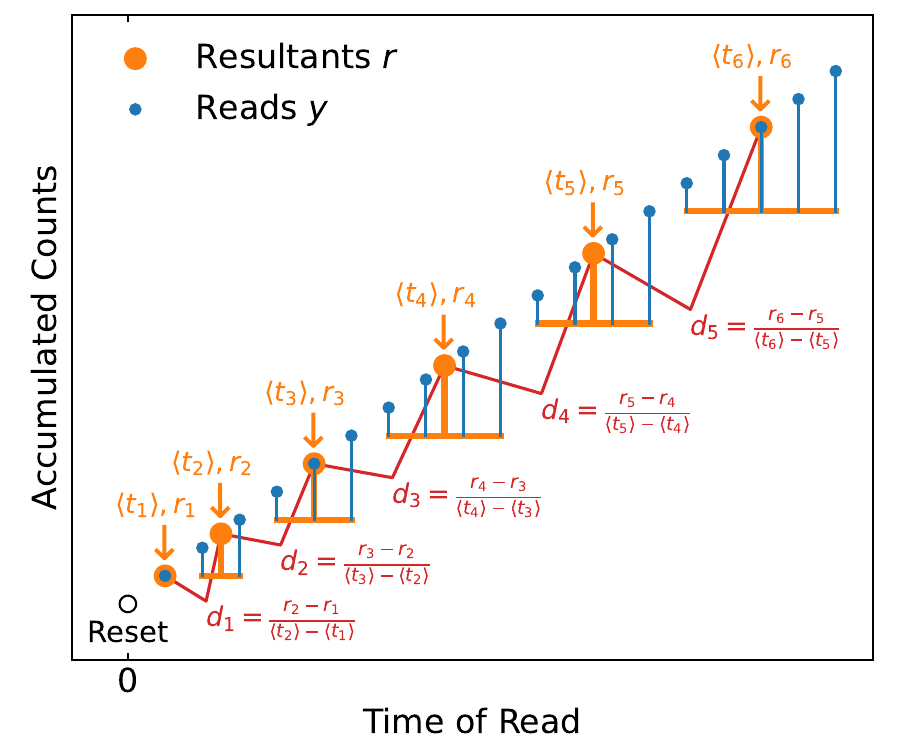}
    \caption{Cartoon showing a ramp consisting of 19 reads (blue points) following an unmeasured reset (black circle); the reads are grouped into six resultants (indicated by orange points).  The resultants and times are defined by Equations \eqref{eq:value_resultant} and \eqref{eq:time_resultant}, respectively.  The red labels indicate the five resultant differences to be used in the algorithms derived in this paper. 
 \label{fig:ramp_cartoon}}
\end{figure}

The first task in fitting a ramp is to derive a covariance matrix for the groups of reads or, in this case, for the differences between sequential groups of reads.  I will denote individual reads by $y$; $N$ reads may be averaged together into a group or resultant that I will denote by $r$:
\begin{equation}
    r_i = \frac{1}{N_i} \sum {\bf y}_i \label{eq:value_resultant}
\end{equation}
where ${\bf y}_i$ refers to the $N_i$ reads that were averaged together to produce resultant $r_i$.  The individual reads $k$ were taken at a series of times $t_k$ measured from the last reset at $t=0$; each read $k$ represents $t_k$ worth of accumulated signal.  The mean time of resultant $i$ is then 
\begin{equation}
    \langle t_i \rangle = \frac{1}{N_i} \sum {\bf t}_i \label{eq:time_resultant}
\end{equation}
where ${\bf t}_i$ are the times since reset of the $N_i$ reads that comprise resultant $i$.  Figure \ref{fig:ramp_cartoon} shows this notation on an example ramp that begins with an unmeasured reset at $t=0$ and consists of 19 reads averaged together into six resultants.

Throughout the rest of this paper I operate almost exclusively in the space of resultant differences.  This serves to make the covariance matrix as close to diagonal as possible.  A photon present in the first read will be present in all subsequent reads, so the covariance matrix of the accumulated counts will have all nonzero elements.  In contrast, nonoverlapping differences between resultants will not share any photons.  The covariance matrix due to read noise is already diagonal in the accumulated counts.  It is not diagonal, but rather tridiagonal, in the space of resultant differences: pairs of resultant differences (e.g.~resultant 5 minus resultant 4 and resultant 2 minus resultant 1) will not share any reads unless the resultant differences are sequential.  Most elements of the covariance matrix will then be zero, and this fact enables the algorithms described in this paper.  

Throughout this paper I will refer to the average of a group of reads as a resultant.  I will assume that I have many resultants $\{N_1, \ldots, N_{n+1}\}$: the first resultant is the unweighted average of $N_1$ reads, etc.  Appendix \ref{sec:intraresultantweights} shows how to treat the case of a weighted average of reads and demonstrates the potential improvement in performance.  I assume $n+1$ resultants so that there are $n$ differences between adjacent resultants; this will make the notation more convenient later.  In Figure \ref{fig:ramp_cartoon}, with six resultants, $n=5$.  The normalized difference between two successive resultants $i+1$ and $i$ is then
\begin{equation}
d_i = \frac{r_{i + 1} - r_i}{\langle t_{i + 1} \rangle 
 - \langle t_i \rangle}
\label{eq:diff}
\end{equation}
where $r_i$ is given by Equation \eqref{eq:value_resultant} $\langle t_i \rangle$ is given by Equation \eqref{eq:time_resultant}; these differences are indicated in red in Figure \ref{fig:ramp_cartoon}.  The quantity $d_i$ has units of counts per unit time.  I will assume henceforth that the counts are in units of electrons.  I will also assume an ideal detector that steadily accumulates counts after the last reset and that is subject only to read noise and photon noise.  
In this section I will first derive the variance and covariance of resultants, and then transform these into the variance and covariance of resultant differences.  The derivation of the covariance matrix that I present is similar to those in \cite{Kubik+Barbier+Castera+etal_2015} and in \cite{Casertano_2022}.  

The variance of a read due to read noise is $\sigma^2$, where $\sigma^2$ is the single read variance (one-half the correlated double sampling variance).  The variance of a resultant due to read noise is then
\begin{equation}
    {\rm Var}(r_i) = \frac{\sigma^2}{N_i} \label{eq:readnoisevar}
\end{equation}
where $N_i$ is the number of reads in resultant $i$.  The covariance between different resultants due to read noise is zero since they do not share any reads.

The covariance between two reads due to photon noise is the expected number of photons that are shared between the two reads, i.e., 
\begin{equation}
    {\rm Cov} (y_i, y_j) = a\cdot {\rm min}(t_i, t_j) 
\end{equation}
for a count rate $a$.  For two different resultants, assuming $i < j$ and that all reads in resultant $i$ precede the first read in resultant $j$, the covariance is given by 
\begin{align}
    {\rm Cov} (r_i, r_j) &= \frac{a}{N_i N_j} \sum_{k = 1}^{N_i} \sum_{l = 1}^{N_j} {\rm min}(t_{i,k}, t_{j,l}) \nonumber \\
    &= \frac{a}{N_i N_j} \sum_{k = 1}^{N_i} \sum_{l = 1}^{N_j} t_{i,k} \nonumber \\
    &= \frac{a}{N_i} \sum_{k = 1}^{N_i} t_{i,k}  \nonumber \\
    &= a \langle t_i \rangle .\label{eq:covar_rirj}
\end{align}
The variance of a single resultant due to photon noise is given by 
\begin{align}
    {\rm Var} (r_i) &= \frac{a}{N_i^2} \sum_{k = 1}^{N_i} \sum_{l = 1}^{N_i} {\rm min}(t_{i,k}, t_{i,l}) . \label{eq:var_singleres_phnoise}
\end{align}
The time of the first read will appear $2 N_i - 1$ times in this double sum, $N_i$ times for each sum minus one from double counting.  The time of the second read will appear $2 N_i - 3$ times, and so on.  The variance can then be written
\begin{align}
    {\rm Var} (r_i) &= \frac{a}{N_i^2} \sum_{k = 1}^{N_i} (2N_i - 2k + 1) t_k .
    \label{eq:var_ri}
\end{align}
Following \cite{Casertano_2022} I define a variance-weighted time $\tau_i$ for each resultant $i$
\begin{align}
    \tau_i = \frac{1}{N_i^2} \sum_{k = 1}^{N_i} (2N_i - 2k + 1) t_k 
\end{align}
so that
\begin{align}
    {\rm Var} (r_i) &= a \tau_i.
\end{align}
If there are a large number $N$ of evenly spaced reads in each resultant,
\begin{equation}
    t_k = t_1 + (k - 1) \frac{\Delta t}{N}
\end{equation}
with the total duration of the resultant being
\begin{equation}
    \Delta t = t_N - t_1 ,
\end{equation}
we have
\begin{align}
    \tau_i &\approx \langle t_i \rangle - 2 \sum_{k = 1}^{N_i} k t_k
\end{align}
and
\begin{align}
    \lim_{N \rightarrow \infty} \tau_i &= \langle t_i \rangle - \frac{\Delta t}{6} .
\end{align}

Using Equations \eqref{eq:covar_rirj} and \eqref{eq:var_ri}, we can now write the variance of the resultant difference $r_{i + 1} - r_{i}$ and the covariance of resultant differences $r_{i + 1} - r_{i}$ and $r_{j + 1} - r_{j}$, including both read noise and photon noise.  For the variance, we have
\begin{align}
    {\rm Var} (r_{i+1}-r_{i}) &= {\rm Var} (r_i) + {\rm Var} (r_{i+1}) - 2\cdot{\rm Cov} (r_i, r_{i+1}) \nonumber \\
    &= \sigma^2 \left(\frac{1}{N_i} + \frac{1}{N_{i+1}}\right) + a \left(\tau_i + \tau_{i+1} - 2 \langle t_i \rangle \right) .
\end{align}
If the resultants each last for a time $\Delta t$, consist of many reads, and occur immediately after one another, the variance becomes
\begin{align}
    {\rm Var} (r_{i+1}-r_{i}) &\approx \sigma^2 \left(\frac{1}{N_i} + \frac{1}{N_{i+1}}\right) + a \left(\frac{2\Delta t}{3}\right) .
\end{align}
This is slightly less than the variance from two reads evenly spaced by $\Delta t$, which would have a factor of unity in place of $\frac{2}{3}$.

For the covariance, with $j=i+1$ (i.e.~consecutive resultant differences), we have
\begin{align}
    {\rm Cov} (r_{j+1}-r_{j}, r_{i+1}-r_{i}) &= {\rm Cov} (r_{i+2}, r_{i+1}) - {\rm Var} (r_{i+1}) - {\rm Cov} (r_{i+2}, r_{i}) + {\rm Cov} (r_{i+1}, r_{i}) \nonumber \\
    &= a \langle t_{i+1} \rangle - \frac{\sigma^2}{N_{i+1}} - a \tau_{i+1} - a \langle t_{i} \rangle + a \langle t_{i} \rangle \nonumber \\
    &= - \frac{\sigma^2}{N_{i+1}} + a \left(\langle t_{i+1} \rangle - \tau_{i+1}\right) .
\end{align}
If the resultants consist of uninterrupted sequences of many reads with no gaps between resultants, and each lasts for $\Delta t$, this covariance becomes 
\begin{align}
    {\rm Cov} (r_{j+1}-r_{j}, r_{i+1}-r_{i}) &\approx - \frac{\sigma^2}{N_{i+1}} + a \left(\frac{\Delta t}{6}\right) . \label{eq:covar_resultants_approx}
\end{align}
The second term would be absent from Equation \eqref{eq:covar_resultants_approx} for single read resultants because the time intervals of the two resultant differences would be fully disjoint and no photons would be shared.  
If $j>i+1$, we have
\begin{align}
    {\rm Cov} (r_{j+1}-r_{j}, r_{i+1}-r_{i}) &= {\rm Cov} (r_{j+1}, r_{i+1}) - {\rm Cov} (r_{j+1}, r_{i}) + {\rm Cov} (r_{j}, r_{i+1}) -  {\rm Cov} (r_j, r_{i}) \nonumber \\
    &= a \langle t_{i+1} \rangle - a \langle t_{i} \rangle + a \langle t_{i + 1} \rangle - a \langle t_{i} \rangle \nonumber \\
    &= 0 \label{eq:zerocov_nonadjacent}.
\end{align}
In other words, only adjacent resultant differences--those that share a resultant--have nonzero covariance. 
For notational convenience I will define
\begin{equation}
    \delta_i t = \langle t_{i + 1}\rangle - \langle t_i \rangle
\end{equation}
so that $\delta_i t$ is the characteristic difference of the integration times in the resultant difference $d_i$.  The scaled resultant differences (Equation \eqref{eq:diff}) are then
\begin{align}
    d_i = \frac{r_{i+1} - r_i}{\delta_i t} .
\end{align}

The covariance matrix of all of the scaled resultant differences $d_i$ may be written as a matrix ${\bf C}_r$ to be multiplied by the read noise variance $\sigma^2$ and a second matrix ${\bf C}_\gamma$ to be multiplied by the photon count rate $a$:
\begin{equation}
    {\bf C} = a {\bf C}_\gamma + \sigma^2 {\bf C}_r.
    \label{eq:covar_sum}
\end{equation}
The read noise matrix has components
\begin{align}
    \left({\bf C}_r\right)_{ij} = 
   \frac{1}{\left( \delta_i t \right) \left( \delta_j t \right)} \times 
 \begin{cases}
    1/N_i + 1/N_{i + 1} & i = j \\
    -1/N_j & {j = i + 1} \\
    -1/N_i & {i = j + 1} \\
 0 & {|i - j|>1}
    \end{cases}
    \label{eq:readnoise}
\end{align}
and the photon noise matrix has components
\begin{align}
    \left({\bf C}_\gamma\right)_{ij} = 
   \frac{1}{\left( \delta_i t \right) \left( \delta_j t \right)} \times 
 \begin{cases}
    \tau_i + \tau_{i+1} - 2\langle t_i \rangle & i = j \\
    \langle t_{j}\rangle - \tau_{j} & {j = i + 1} \\
    \langle t_{i}\rangle - \tau_{i} & {i = j + 1} \\
 0 & {|i - j|>1}
    \end{cases}
    \label{eq:cmat_readnoise}
\end{align}

In Equation \eqref{eq:covar_sum}, ${\bf C}_\gamma$ and ${\bf C}_r$ depend only on the properties of the readout pattern, i.e., the number and times of the individual reads within each resultant: they do not need to be computed separately for every pixel.  Both are tridiagonal, so the total covariance matrix ${\bf C}$ will also be tridiagonal.  This fact was also pointed out by \cite{Kubik+Barbier+Castera+etal_2015}.

The arguments and derivations above provide the elements of the tridiagonal covariance matrix of the resultant differences ${\bf d} = \{d_i\}$ as
\begin{equation}
    {\bf C} = 
    \begin{bmatrix}
    \alpha_1 & \beta_1 & 0 & 0 & \ldots \\
    \beta_1 & \alpha_2 & \beta_2 & 0 & \ldots \\
    0 & \ddots & \ddots & \ddots \\
    0 & \ldots & \beta_{n-2} & \alpha_{n-1} & \beta_{n-1} \\
    0 & \ldots & 0 & \beta_{n-1} & \alpha_n \\
    \end{bmatrix}. \label{eq:C_tridiag}
\end{equation}
Each element $\alpha$ and $\beta$ is the sum of a term scaled by a given pixel's photon rate $a$ and another term scaled by the read noise variance $\sigma^2$:
\begin{align}
    \alpha_i &= \sigma^2 \left(\frac{1}{\left(\delta_i t\right)^2}\right) \left(\frac{1}{N_i} + \frac{1}{N_{i+1}}\right) + a \left( \frac{1}{\left(\delta_i t\right)^2} \right) \left( \tau_i + \tau_{i+1} - 2\langle t_i \rangle \right) \label{eq:alphas} \\
    \beta_i &= \sigma^2 \left(\frac{1}{\left(\delta_i t\right)\left(\delta_{i+1} t\right)}\right) \left(\frac{-1}{N_{i+1}}\right) + a \left( \frac{1}{\left(\delta_i t\right)\left(\delta_{i+1} t\right)} \right) \left( \langle t_{i+1} \rangle - \tau_{i+1} \right)  \label{eq:betas}
\end{align}
The read noise $\sigma$ may typically be measured for each pixel, but the true count rate $a$ will be unknown.  For the following section I will assume that the count rate is given and will derive the slope of the best-fit ramp, its uncertainty, and its goodness-of-fit $\chi^2$.  I will then turn to the problem of estimating the covariance matrix itself.

\subsection{Including the Reset Value} \label{sec:covar_reset}

The preceding discussion derived the covariance matrix for the differences of adjacent resultants.  For some applications the reset value is also useful.  This could be for applying a nonlinearity correction, for monitoring the detector stability, or even for using the first read to measure the count rate.  The precision of measuring the count rate using the first resultant alone is limited by $kTC$ noise in the reset value.

If we wish to include the reset value, then we will also make use of the first resultant $r_1$.  We define $d_0$ as
\begin{equation}
    d_0 \equiv \frac{r_1}{\langle t_1 \rangle} \label{eq:d0}
\end{equation}
so that, in the absence of noise,
\begin{equation}
    d_0 = a + \frac{b}{\langle t_1 \rangle}
\end{equation}
where $b$ is the reset value (the counts in a pixel at $t=0$).  If we wish to measure $b$, we can prepend $d_0$ to the vector $\left\{d_1, \ldots, d_n\right\}$.  We also need to prepend values to both $\alpha$ and $\beta$ for the covariance matrix.  The value of $\alpha_0$ will be
\begin{equation}
    \alpha_0 = \frac{1}{\langle t_1 \rangle^2}{\rm Var}\left(r_1\right) = \sigma^2 \left(\frac{1}{N\langle t_1 \rangle^2}\right) + a \left(\frac{\tau_1}{\langle t_1 \rangle^2} \right) \label{eq:alpha0}
\end{equation}
while the value of $\beta_0$ will be
\begin{align}
    \beta_0 &= \frac{1}{\langle t_1 \rangle \left(\delta_1 t\right)}{\rm Cov}\left(r_2, r_1\right) - \frac{1}{\langle t_1 \rangle \left(\delta_1 t\right)}{\rm Var}\left(r_1\right) \nonumber \\
    &= \sigma^2 \left(\frac{-1}{N\langle t_1 \rangle \left(\delta_1 t\right)}\right) + a \left( \frac{\langle t_1 \rangle - \tau_1}{\langle t_1 \rangle \left(\delta_1 t\right)} \right). \label{eq:beta0}
\end{align}
Equations \eqref{eq:alphas} and \eqref{eq:betas} may also be used directly if we take $\delta_0 t = \langle t_1 \rangle$ and $1/N_0=0$.  The covariance matrix remains tridiagonal.

\section{Fitting a Ramp} \label{sec:fitramp}

With the covariance matrix defined by Equation \eqref{eq:C_tridiag} via Equations \eqref{eq:alphas} and \eqref{eq:betas}, we want to fit the scaled resultant differences.  I will defer the calculation including the reset value, which uses the additional elements of the covariance matrix given in Section \ref{sec:covar_reset}, for Section \ref{sec:fit_reset}.

All scaled resultant differences $d_i$ for $i=1, \ldots n$ should have the same value in the absence of noise assuming the astrophysical count rate to be constant and the detector to be linear and well-behaved.  The likelihood of a model consisting of a single count rate $a$ is then
\begin{equation}
-2 \ln {\cal L} = \chi^2 = \left({\bf d} - a \cdot {\bf 1} \right)^T {\bf C}^{-1} \left({\bf d} - a \cdot {\bf 1} \right)
\end{equation}
where ${\bf 1}$ refers to a vector of all ones.  I can find the maximum likelihood count rate by differentiating this and setting it equal to zero:
\begin{align}
    \frac{d\chi^2}{da} = 0 = 2 \cdot {\bf 1}^T {\bf C}^{-1} \left({\bf d} - a \cdot {\bf 1}\right)
\end{align}
or
\begin{align}
    a = \left( {\bf 1}^T {\bf C}^{-1} {\bf d} \right) \left( {\bf 1}^T {\bf C^{-1}} {\bf 1} \right)^{-1} .
\end{align}
The formula for $\chi^2$ itself may be expanded out as
\begin{equation}
    \chi^2 = \left( {\bf d}^T {\bf C}^{-1} {\bf d} \right) + 2a \left( {\bf 1}^T {\bf C}^{-1} {\bf d} \right) + a^2 \left( {\bf 1}^T {\bf C}^{-1} {\bf 1} \right) .
\end{equation}
These equations all include a matrix inverse and matrix multiplications.  A general matrix inverse has a computational cost of $n^3$ where $n$ is the dimensionality of the matrix, while matrix multiplication with a vector has a cost of $n^2$.  These costs could be unacceptable if there are many reads or many resultants for millions of pixels.  In the following I will show that the best-fit $a$ and $\chi^2$ may be computed using closed formulas for a cost that is linear in the number of resultant differences $n$.

I will begin by computing the inverse of the covariance matrix, using the formula for a tridiagonal matrix.  I will first define some helper variables using recursion relations \citep[Equations (1.1), (1.3) and (1.4) of][]{USMANI1994413}.  I use the same notation as \citeauthor{USMANI1994413} for the helper variables but I adopt Greek letters for the elements of the covariance matrix following Equations \eqref{eq:alphas} and \eqref{eq:betas}.  I have
\begin{align}
    \theta_0 &= 1 \label{eq:theta0} \\
    \theta_1 &= \alpha_1 \\
    \theta_i & = \alpha_i \theta_{i-1} - \beta_{i-1}^2 \theta_{i-2}
\end{align}
and
\begin{align}
    \phi_{n+1} &= 1 \\
    \phi_{n} &= \alpha_n \\
    \phi_i &= \alpha_i \phi_{i+1} - \beta_i^2 \phi_{i+2}. \label{eq:phi_i}
\end{align}
The inverse of the covariance matrix is then given by 
\begin{equation}
    \left({\bf C^{-1}} \right)_{ij} = \begin{cases}
    (-1)^{i+j} \beta_i \cdots \beta_{j-1} \theta_{i-1}\phi_{j+1}/\theta_n & i < j \\
    \theta_{i-1}\phi_{i+1}/\theta_n & i = j \\
    (-1)^{i+j} \beta_j \cdots \beta_{i-1} \theta_{j-1}\phi_{i+1}/\theta_n & i > j
    \end{cases}. \label{eq:tridiag_inverse}
\end{equation}
I will further define
\begin{align}
    B_i &= \prod_{j=1}^{i-1} \beta_j ~~{\rm with}~~ B_1 = 1 \label{eq:B_i} \\
    \Phi_i &= \sum_{j=i+1}^n (-1)^{j} \frac{B_{j}}{B_i} \phi_{j+1} ~~{\rm with}~~ \Phi_{n} = 0 \label{eq:Phi_i} \\
    \Theta_i &= \sum_{j=0}^{i - 1} (-1)^{j+1} \theta_j \frac{B_i}{B_{j+1}} \\
    (\Theta {\rm D})_i &= \sum_{j=1}^{i} (-1)^{j} d_{j} \theta_{j-1} \frac{B_i}{B_{j}} ~~{\rm with}~~ (\Theta {\rm D})_0 = 0 \label{eq:ThetaD_i}
\end{align}
Each of these is computable with a cost linear in the number of resultant differences.  However, they are problematic if any of the $\beta$ terms are zero.  We can avoid this possibility by using the following equivalent recursion relations:
\begin{align}
    \Phi_i &= \beta_i \Phi_{i+1} + (-1)^{i+1} \beta_i \phi_{i+2} \\
    \Theta_i &= \beta_{i-1} \Theta_{i-1} + (-1)^i \theta_{i-1} \\
    \left(\Theta D \right)_i &= \beta_{i-1} \left(\Theta D \right)_{i-1} + (-1)^i d_i \theta_{i-1} 
\end{align}
with the initial conditions
\begin{align}
    \Phi_{n}&=0 \\
    \Theta_1&=-\theta_0 \\
    \left(\Theta D\right)_0 &= 0 \\
    \left(\Theta D\right)_1 &= -d_1\theta_0
\end{align}

With these definitions, I will compute the terms I need to solve.  First, the best-fit slope is given by
\begin{equation}
    a = \left(\sum_{i=1}^n d_i \sum_{j=1}^n \left({\bf C^{-1}}\right)_{ij}\right) \left(\sum_{i=1}^n \sum_{j=1}^n \left({\bf C^{-1}}\right)_{ij}\right)^{-1}.
    \label{eq:ramp_equation}
\end{equation}
The first term in Equation \eqref{eq:ramp_equation} may be written as
\begin{align}
    \sum_{i=1}^n d_i \sum_{j=1}^n \left({\bf C^{-1}}\right)_{ij} &= \sum_{i=1}^n d_i \left( \sum_{j=1}^i \left(-1\right)^{i+j} \frac{B_i \theta_{j-1} \phi_{i+1}}{B_j \theta_n} + \sum_{j=i+1}^n \left(-1\right)^{i+j} \frac{B_j \theta_{i-1} \phi_{j+1}}{B_i \theta_n}\right) \nonumber \\
    &= \sum_{i=1}^n d_i \left( \left(-1\right)^i \frac{\phi_{i+1}}{\theta_n} \Theta_i + \left(-1\right)^i \frac{\theta_{i-1} }{\theta_n} \Phi_i \right)  \nonumber \\
    &= \sum_{i=1}^n d_i \frac{\left(-1\right)^i}{\theta_n} \left(  \phi_{i+1} \Theta_i +  \theta_{i-1} \Phi_i \right) .
\end{align}
The second term will look just like the first term but without the $d$ factor, i.e.,
\begin{align}
    \sum_{i=1}^n \sum_{j=1}^n \left({\bf C^{-1}}\right)_{ij} &= \sum_{i=1}^n  \frac{\left(-1\right)^i}{\theta_n} \left(  \phi_{i+1} \Theta_i +  \theta_{i-1} \Phi_i \right) .
\end{align}
The only term that remains to compute for $\chi^2$ is
\begin{align}
        \sum_{i=1}^n \sum_{j=1}^n d_i d_j \left({\bf C^{-1}}\right)_{ij} .
\end{align}
For this term, I will use the symmetry of the covariance matrix to write
\begin{align}
        \sum_{i=1}^n \sum_{j=1}^n d_i d_j \left({\bf C^{-1}}\right)_{ij} &= 2 \sum_{i=1}^n \sum_{j=1}^{i-1} d_i d_j \left({\bf C^{-1}}\right)_{ij} + \sum_{i=1}^n d_i^2 \left({\bf C^{-1}}\right)_{ii} \nonumber \\
        &= 2 \sum_{i=1}^n d_i \sum_{j=1}^{i-1} d_j \left(-1\right)^{i+j} \frac{B_i \theta_{j-1} \phi_{i+1}}{B_j \theta_n} + \sum_{i=1}^n d_i^2 \frac{\theta_{i-1}\phi_{i+1}}{\theta_n} \nonumber \\
        &= 2 \sum_{i=1}^n d_i \sum_{j=1}^{i-1} d_j \left(-1\right)^{i+j} \frac{\beta_{i-1} B_{i-1} \theta_{j-1} \phi_{i+1}}{B_j \theta_n} + \sum_{i=1}^n d_i^2 \frac{\theta_{i-1}\phi_{i+1}}{\theta_n} \nonumber \\
        &= 2 \sum_{i=1}^n d_i \frac{(-1)^i}{\theta_n} \phi_{i+1}\beta_{i-1} (\Theta{\rm D})_{i-1} + \sum_{i=1}^n d_i^2 \frac{\theta_{i-1}\phi_{i+1}}{\theta_n} 
\end{align}
taking $\beta_0=1$ for the $i=1$ term of the first sum.  Again, this is computable at a cost linear in the number of resultant differences. \\

So, to sum up, I will define
\begin{align}
    {\cal A} &= 2 \sum_{i=1}^n d_i \frac{(-1)^i}{\theta_n} \phi_{i+1}(\Theta{\rm D})_{i-1} + \sum_{i=1}^n d_i^2 \frac{\theta_{i-1}\phi_{i+1}}{\theta_n} \label{eq:calA} \\
    {\cal B} &= \sum_{i=1}^n d_i \frac{\left(-1\right)^i}{\theta_n} \left( \phi_{i+1} \Theta_i +  \theta_{i-1} \Phi_i \right) \label{eq:calB} \\
    {\cal C} &= \sum_{i=1}^n  \frac{\left(-1\right)^i}{\theta_n} \left( \phi_{i+1} \Theta_i +  \theta_{i-1} \Phi_i \right) \label{eq:calC} .
\end{align}
The best-fit count rate is then
\begin{equation}
    a = {\cal B}/{\cal C} ,
\end{equation}
its standard error is
\begin{equation}
    \sigma^2_a = 1/{\cal C},
\end{equation}
and the best-fit $\chi^2$ value is
\begin{align}
    \chi^2_{\rm best} &= {\cal A} - 2a {\cal B} + a^2 {\cal C} \nonumber \\
    &= {\cal A} - \frac{{\cal B}^2}{\cal C} 
    . \label{eq:chisq_abc}
\end{align}

This section showed that I can compute the general up-the-ramp count rate with the full covariance matrix at a cost that is linear in the number of resultant differences.  For a very small additional cost (evaluating the ${\cal A}$ term), I can also compute $\chi^2$ and see whether a constant count rate is a good fit to the data.  There is no need to precompute coefficients or interpolate within different signal-to-noise regimes.  The full covariance matrix will be calculated once per frame as a term that is proportional to the photon rate at a given pixel and a second term that is proportional to the read noise variance at each pixel. 

\subsection{Fitting the Reset Value} 
\label{sec:fit_reset}

If we want to fit for the reset value, we use the tridiagonal covariance matrix with the additional $\alpha_0$ and $\beta_0$ defined by Equations \eqref{eq:alpha0} and \eqref{eq:beta0}, and the additional scaled resultant defined by Equation \eqref{eq:d0}.  The equation for $\chi^2$ becomes
\begin{align}
    \chi^2 = \left({\bf d} - a \cdot {\bf 1} - \frac{b}{\langle t_1 \rangle} \cdot {\bf i} \right)^T {\bf C}^{-1} \left({\bf d} - a \cdot {\bf 1}  - \frac{b}{\langle t_1 \rangle} \cdot {\bf i} \right)
\end{align}
where ${\bf i}$ is a vector that is one in the first entry and zero elsewhere.  This may be expanded to obtain
\begin{align}
    \chi^2 = {\bf d}^T {\bf C}^{-1} {\bf d} + a^2 \left( {\bf 1}^T {\bf C}^{-1} {\bf 1}\right) + \frac{b^2}{\langle t_1 \rangle^2} \left( {\bf i}^T {\bf C}^{-1} {\bf i} \right) - 2 a \left( {\bf 1}^T {\bf C}^{-1} {\bf d}\right) - 2 \frac{b}{\langle t_1 \rangle} \left( {\bf i}^T {\bf C}^{-1} {\bf d}\right) + 2 a\frac{b}{\langle t_1 \rangle} \left( {\bf i}^T {\bf C}^{-1} {\bf 1}\right). \label{eq:chisq_withreset_expanded}
\end{align}
Some of these terms were already computed in the first part of Section \ref{sec:fitramp}, allowing me to write
\begin{align}
    \chi^2 = {\cal A} + a^2 {\cal C} + \frac{b^2}{\langle t_1 \rangle^2} \left( {\bf i}^T {\bf C}^{-1} {\bf i} \right) - 2 a {\cal B} - 2 \frac{b}{\langle t_1 \rangle} \left( {\bf i}^T {\bf C}^{-1} {\bf d}\right) + 2 a\frac{b}{\langle t_1 \rangle} \left( {\bf i}^T {\bf C}^{-1} {\bf 1}\right). \label{eq:chisq_reset_intermediate}
\end{align}
The term ${\bf i}^T {\bf C}^{-1} {\bf i}$ is given in Equation \eqref{eq:tridiag_inverse} as
\begin{equation}
    {\bf i}^T {\bf C}^{-1} {\bf i} = C^{-1}_{11} = \frac{\theta_0 \phi_2}{\theta_n}
\end{equation}
while ${\bf i}^T {\bf C}^{-1} {\bf 1}$ may be written using just the first term in the sum of Equation \eqref{eq:calC}:
\begin{align}
    {\bf i}^T {\bf C}^{-1} {\bf 1} \equiv {\cal C}'_1 = \frac{-1}{\theta_n} \left(\phi_2 \Theta_1 + \theta_0 \Phi_1 \right) .
\end{align}
In all of these formulas the $\beta$ and $\alpha$ values prepended to the arrays in Equations \eqref{eq:alphas} and \eqref{eq:betas} are indexed starting at 1.  In other words, where $\beta_1$ appears in these equations, it now refers to the value in Equation \eqref{eq:beta0}, and where $d_1$ appears, it refers to the value in Equation \eqref{eq:d0}.

To write the remaining term in Equation \eqref{eq:chisq_reset_intermediate} more conveniently, I will define one additional quantity 
\begin{equation}
    \left(\Phi D\right)_j = \sum_{i=j + 1}^n (-1)^i d_i \phi_{i+1} \frac{B_i}{B_j}  ~~{\rm with} ~~ \left(\Phi D\right)_{n-1}=0 .
\end{equation}
As for the terms in Equations \eqref{eq:Phi_i}--\eqref{eq:ThetaD_i}, this is equivalently defined by the recursion relation
\begin{align}
    \left(\Phi D\right)_{n-1} &= 0 \\
    \left(\Phi D\right)_{j} &= \beta_j \left(\Phi D\right)_{j+1} + \left(-1\right)^{j+1} \beta_j d_{j+1}\phi_{j+2}
\end{align}
which avoids the possibility of division by zero.  With this definition, I can write
\begin{equation}
    {\bf d}^T {\bf C}^{-1} {\bf i} \equiv {\cal B}'_1 = \frac{-1}{\theta_n} \left( \phi_{2} (\Theta D)_1 +  \theta_{0} (\Phi D)_1 \right) 
\end{equation}
and finally
\begin{equation}
    \chi^2 = {\cal A} + a^2 {\cal C} + \frac{b^2}{\langle t_1 \rangle^2} C^{-1}_{11} - 2\frac{b}{\langle t_1 \rangle} {\cal B}'_1 - 2a {\cal B} + 2a \frac{b}{\langle t_1 \rangle} {\cal C}'_1.
    \label{eq:chi2_resetval}
\end{equation}
In some cases, there may be a prior placed on the reset value $b$.  If the reset value is stable up to $kTC$ noise and only the first resultant is usable, then the use of a prior on $b$ is the only way to obtain a constraint on the count rate $a$.  Assuming a Gaussian prior with a mean of $z$ and an uncertainty $\sigma_z$, the expression for $\chi^2$ becomes
\begin{equation}
    \chi^2 = {\cal A} + a^2 {\cal C} + \frac{b^2}{\langle t_1 \rangle^2} C^{-1}_{11} - 2\frac{b}{\langle t_1 \rangle} {\cal B}'_1 - 2a {\cal B} + 2a \frac{b}{\langle t_1 \rangle} {\cal C}'_1 + \frac{\left(b - z\right)^2}{\sigma_z^2}.
    \label{eq:chi2_resetval_withprior}
\end{equation}
Setting $z=0$ and $\sigma_z=\infty$ recovers the case of a uniform prior on the reset value.

The first step to computing the best $\chi^2$ is to differentiate $\chi^2$ and set the result equal to zero:
\begin{align}
    \frac{\partial \chi^2}{\partial a} &= 0 = 2a {\cal C} - 2 {\cal B} + 2\frac{b}{\langle t_1 \rangle} {\cal C}'_1 \\
    \frac{\partial \chi^2}{\partial b} &= 0 = 2\frac{b}{\langle t_1 \rangle^2} C^{-1}_{11} - 2 \frac{{\cal B}'_1}{\langle t_1 \rangle} + 2\frac{a}{\langle t_1 \rangle} {\cal C}'_1 + 2\left(\frac{b - z}{\sigma_z^2}\right) .
\end{align}
This yields
\begin{align}
    b &= \left(\frac{{\cal B}'_1}{\langle t_1 \rangle} - \frac{{\cal C}'_1 {\cal B}}{{\cal C} \langle t_1 \rangle} + \frac{z}{\sigma^2_z}\right) \left(\frac{C_{11}^{-1}}{\langle t_1 \rangle^2} - \frac{{\cal C}_1'^2}{{\cal C}\langle t_1 \rangle^2} + \frac{1}{\sigma_z^2}\right)^{-1} \\
    a &= \frac{\cal B}{\cal C} - b\left(\frac{{\cal C}'_1}{{\cal C}\langle t_1 \rangle} \right) .
\end{align}

The covariance matrix for $a$ and $b$ is then
\begin{align}
    {\bf C}^{-1}(a, b) &= 
    \begin{bmatrix}
        \frac{1}{2}\frac{\partial^2 \chi^2}{\partial a^2} & \frac{1}{2}\frac{\partial^2 \chi^2}{\partial a \partial b} \\
        \frac{1}{2}\frac{\partial^2 \chi^2}{\partial a \partial b} & \frac{1}{2}\frac{\partial^2 \chi^2}{\partial b^2}
    \end{bmatrix} \nonumber \\
    &= \begin{bmatrix}
        {\cal C} & {\cal C}'_1/\langle t_1 \rangle \\
        {\cal C}'_1/\langle t_1 \rangle & C^{-1}_{11}/\langle t_1 \rangle^2 + 1/\sigma_z^2
    \end{bmatrix} .
\end{align}
This may be inverted by hand to get the standard errors on $a$ and $b$ and their covariance.

\subsection{Omitting One or More Resultant Differences} \label{subsec:omitreads}

Sometimes a resultant difference is corrupted, e.g., by a cosmic ray: there can be a jump in counts between two resultants.  There can also be a jump within a resultant, in which case two resultant differences must be discarded (both of the differences that contain the resultant with a jump).  Saturation of a pixel or of its neighbor can also corrupt all resultant differences after the onset of saturation.

One way of discarding a resultant difference is to write down a new, smaller covariance matrix for the pixel in question that omits the corrupted difference(s).  In order to facilitate the use of the equations developed in this section, 
I adopt a different approach.  Assuming that we wish to discard resultant difference $j$, i.e.~$d_j$, I first decouple $d_j$ from the other differences by setting $\beta_{j-1}=\beta_j=0$.  This renders the covariance matrix block diagonal.  The inverse of the covariance matrix now has elements in row/column $j$
\begin{equation}
    C^{-1}_{jk} = C^{-1}_{kj} = \frac{1}{\alpha_j} \delta_{jk}
\end{equation}
where $\delta_{jk}$ is the Kronecker delta.  We need to set these to zero or to ensure that terms containing them are zero.  The elements within the sum for ${\cal C}$ in Equation \eqref{eq:calC} are the column-summed elements of ${\bf C}^{-1}$; so we can set the $j$ term to zero (equivalently, we can set $\Theta_j=\Phi_j=0$).  For ${\cal B}$ and ${\cal A}$, the only resultant difference that multiplies the $j$ row or column of ${\bf C}^{-1}$ is $d_j$, so we can set $d_j = 0$.  

In sum, if we wish to ignore resultant difference $j$, we set 
\begin{equation}
\beta_{j-1}=\beta_j=\Theta_j=\Phi_j=d_j=0.
\end{equation}
We can do this for any number of resultant differences in any subset of pixels, and continue to apply the equations derived above.  This statement holds true whether or not we are fitting for the reset value.

\section{Biases and Estimating the Covariance Matrix} \label{sec:bias}

Sections \ref{sec:covmatrix} and \ref{sec:fitramp} assumed that the covariance matrix is known.  In general the read noise of each pixel may be accurately known, but that pixel's true count rate will not be known.  The covariance matrix must first be approximated using the resultants themselves.  This could introduce biases.  In this section I will compute those biases to first order and show that fitting for the count rate using two iterations effectively avoids them.  Prior to this, I will treat the case of discretely varying weighting schemes to estimate the count rate, as presented by \cite{Fixsen+Offenberg+Hanisch+etal_2000} and refined by \cite{Casertano_2022}, showing that it is also biased and deriving an analytic formula for the bias.

\subsection{Biases with Approximate Weights: The Discrete Case} \label{appendix:bias}

The ramp-fitting approach of \cite{Fixsen+Offenberg+Hanisch+etal_2000}, \cite{Offenberg+Fixsen+Rauscher+etal_2001}, and \cite{Casertano_2022} uses fixed weights for the different resultants, with the weights determined by the signal-to-noise ratio as estimated by the difference between the first and last resultants.  The use of discrete weights does introduce biases in the recovered count rate near the signal-to-noise ratios at which the weights are discontinuous.  This section provides intuition for the source of the bias and then presents a calculation of its magnitude as a function of the read noise, the true count rate, and the readout pattern.  

A bias exists because the estimated count rate is a weighted sum of the resultants, but this is covariant with the difference between the first and the last resultants which is used to determine the weights.  I will use $s$ to denote the difference between the last and first resultants.  The signal-to-noise ratio estimate used by \cite{Casertano_2022} is
\begin{equation}
    {\rm SNR} = {\rm max} \left(0, \frac{s}{\sqrt{s + \sigma^2}} \right)
\end{equation}
where $\sigma^2$ is the read noise.  
The estimate of the count rate is a weighted sum of the resultants,
\begin{equation}
    a = \sum_i w_i r_i . \label{eq:weights_fixsen}
\end{equation}
Different sets of weights are used depending on the signal-to-noise ratio inferred from $s$.  The inferred count rate $a$ will be covariant with $s$.  As the signal-to-noise ratio increases, the first and last resultants are weighted more heavily in the sum of Equation \eqref{eq:weights_fixsen} and this covariance becomes stronger.  Near a break in the weighting scheme the joint distribution of $a$ and $s$ then has a discontinuity along a line of constant $s$: the covariance above this line is larger than the covariance below this line.  A joint distribution that is symmetric far from any discontinuity in the weighting scheme becomes asymmetric near a discontinuity.

\begin{figure*}
\includegraphics[width=\textwidth]{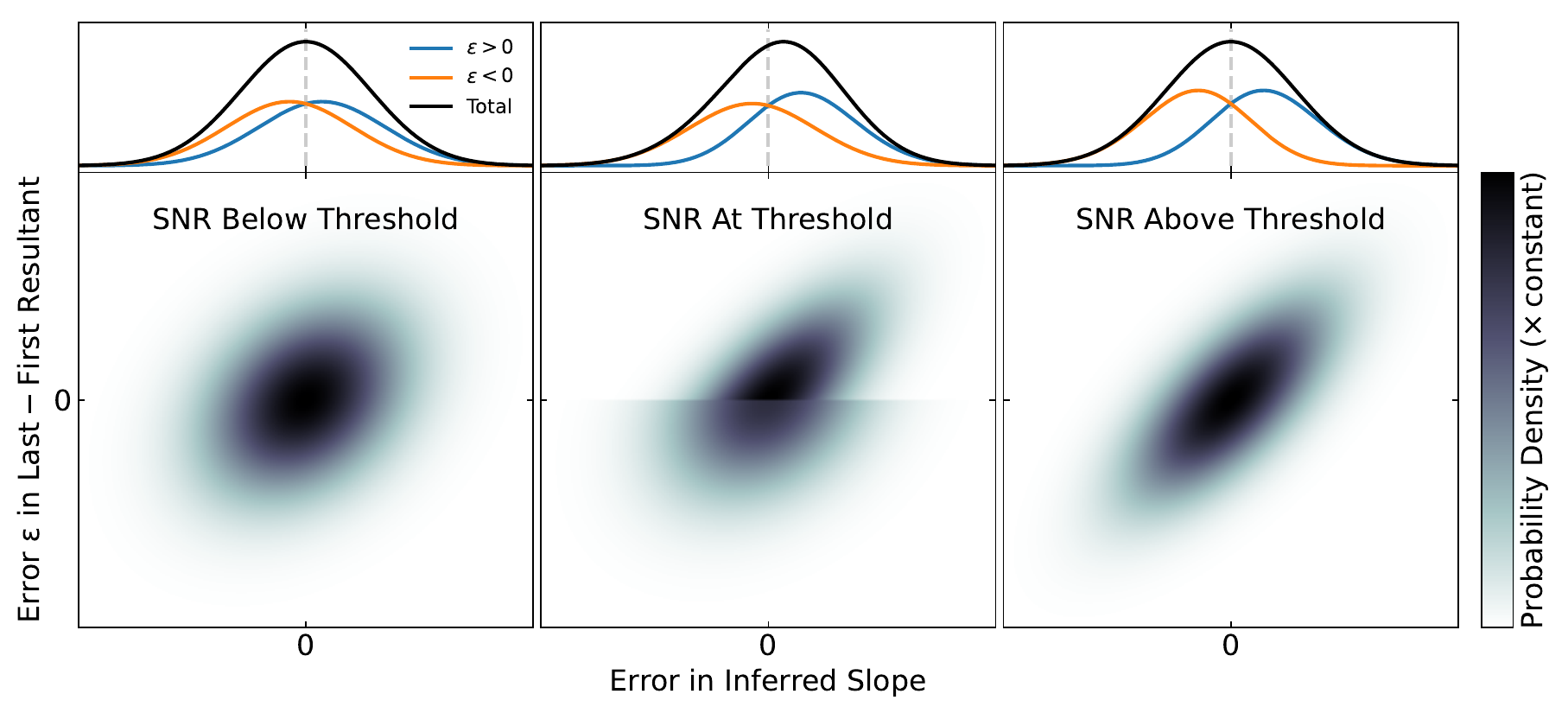}
\caption{Illustration of the phenomenon that leads to bias when using weights that vary discontinuously to estimate the count rate. 
 The weights used depend on the difference between the first and last resultant ($y$-axis). 
 The error in this quantity is covariant with the error in the inferred slope. 
 When the weights used are independent of the count rate, this covariance leads to an error ellipse and an expectation value of zero for the error in the inferred count rate.  Near a threshold between two weighting schemes, however, two different Gaussians are combined, and the expectation of the error in the inferred slope can be nonzero. The top panels show the probability densities marginalized over the error in the difference between the first and last resultant, decomposed by the sign of this error.
 \label{fig:bias_demo_discrete}}
\end{figure*}

Figure \ref{fig:bias_demo_discrete} illustrates the idea expressed above.  The figure shows two different Gaussians, each with the same uncertainty in the inferred slope and in the difference between the first and the last read, but with different covariances between these two quantities.  The left and right panels show each two-dimensional Gaussian individually; these would correspond to two different sets of weights.  Each Gaussian shows the joint probability density of realizing a value of the fitted slope and of the difference between the first and last resultants.  The middle panel shows what would happen at a discontinuity in the weights: the two-dimensional Gaussians differ at a threshold in the error in the last minus the first resultant.  If the error in the last minus the first resultant is positive, the estimated S/N is slightly higher than the true S/N, and the weights corresponding to the right Gaussian are used.  If the error is negative, the S/N is slightly underestimated, and the weights corresponding to the left Gaussian are used.

Far from a discontinuity in the weights, in the left and right panels, the joint distribution of $a$ and $s$ is symmetric and its center-of-mass is at zero error in both directions.  The probability densities marginalized over the error in $s$ are symmetric when the marginalization is restricted to positive or negative errors in $s$; this is shown by the blue and orange lines in the top panels.  Near a discontinuity, however, these symmetries no longer hold.  The center of mass of the two-dimensional distribution is at zero error in $s$ (because this uncertainty remains symmetric), but it is no longer at zero error in $a$.  The larger covariance at larger measured values of $s$ leads to an expectation value of the error in $a$ that is greater than zero, i.e., a positive bias.  The probability density marginalized over the error in $s$ is no longer symmetric when restricted to either positive or negative errors in $s$, and the total marginalized distribution has a positive mean (black line, top middle panel).

To calculate the bias of the discrete weighting scheme, we first need the variance of $s$, the variance of $a$, and the covariance of $s$ and $a$.  The variance of $a$ is given in \cite{Casertano_2022} while the variance of $s$ is given in Section \ref{sec:covmatrix}.  Their covariance is given by
\begin{align}
    {\rm Cov}(s, a) &= \sum_i w_i \left( {\rm Cov}(r_{n + 1}, r_i) - {\rm Cov}(r_{1}, r_i) \right) \nonumber \\
    &= w_{n+1} \left( {\rm Var}\left(r_{n+1}\right) - {\rm Cov}\left(r_{n + 1}, r_1 \right) \right) + w_{1} \left( {\rm Cov}\left(r_{n + 1}, r_1 \right) - {\rm Var}\left(r_{1}\right)\right) + \sum_{i=2}^n w_i \left( {\rm Cov}(r_{n + 1}, r_i) - {\rm Cov}(r_{1}, r_i) \right)  \nonumber \\
    &= w_{n+1} \left(a \tau_{n+1} - a \langle t_1 \rangle + \sigma^2 \right) + w_{1} \left(  a \langle t_1 \rangle - a \tau_1 - \sigma^2\right) + \sum_{i=2}^n a w_i \left( \langle t_i \rangle -  \langle t_1 \rangle\right) \label{eq:cov_a_s} 
\end{align}
where $r_1$ and $r_{n+1}$ denote the first and last resultant, respectively.  The joint distribution between $s$ and $a$ is then given by the covariance matrix
\begin{equation}
    {\bf \Sigma} = \begin{bmatrix}
        \sigma^2_s & {\rm Cov}(s, a) \\
        {\rm Cov}(s, a) & \sigma^2_a
    \end{bmatrix} .
\end{equation}
I will denote the elements of the inverse of this matrix as
\begin{equation}
    {\bf \Sigma}^{-1} = \begin{bmatrix}
        m_{11} & m_{12} \\
        m_{21} & m_{22}
    \end{bmatrix} 
\end{equation}
with $m_{12} = m_{21}$.  The covariance matrix will differ above and below any threshold in $s$; I will assume that there is a discontinuity at $s=\lambda$, and that the elements of ${\bf \Sigma}^{-1}$ above and below $\lambda$ are denoted by $m$ and $m'$, respectively.  In this case the expectation value of the error on $a$, $a - \tilde{a}$, is
\begin{align}
    \langle a - \tilde{a} \rangle = &\int_{-\infty}^\infty a\, da \int_{-\infty}^\lambda \frac{ds}{2\pi \sqrt{\rm det \bf \Sigma'}} \exp \left( -\frac{1}{2} \left( m'_{11} s^2 + m'_{22} a^2 + 2 m'_{12} s a \right) \right) \nonumber \\
    & \qquad + \int_{-\infty}^\infty a\, da \int_\lambda^\infty \frac{ds}{2\pi \sqrt{\rm det \bf \Sigma}} \exp \left( -\frac{1}{2} \left( m_{11} s^2 + m_{22} a^2 + 2 m_{12} s a \right) \right).
\end{align}
I will focus only on the second term, which I will denote $\langle a - \tilde{a}\rangle_+$.  The first term may be equivalently written with limits on $s$ from $-\lambda$ to $\infty$ by replacing $s$ with $-s$.  I first complete the square and integrate over $a$:
\begin{align}
    \langle a - \tilde{a} \rangle_+ = \int_\lambda^\infty \frac{ds}{2\pi \sqrt{\rm det \bf \Sigma}} \int_{-\infty}^\infty \left( a + \frac{m_{12}}{m_{22}} s - \frac{m_{12}}{m_{22}} s \right) \, da \exp \left( -\frac{m_{22}}{2} \left(a + m_{12}/m_{22} s\right)^2 - \frac{s^2}{2} \left( m_{11} - \frac{m_{12}^2}{m_{22}} \right) \right) .
\end{align}
The odd portion of this function integrates to zero, while the even portion is an ordinary Gaussian integral.  I will also use the facts that
\begin{align}
    \frac{m_{22}}{m_{11}m_{22} - m_{12}^2} &= \sigma^2_s \\
    \frac{-m_{12}}{m_{11}m_{22} - m_{12}^2} &= {\rm Cov}(s, a) \\
    \frac{1}{m_{11}m_{22} - m_{12}^2} &= {\rm det} {\bf \Sigma} .
\end{align}
This allows me to write
\begin{align}
    \langle a - \tilde{a} \rangle_+ &= \int_\lambda^\infty \frac{s\, ds}{2\pi \sqrt{\rm det \bf \Sigma}} \int_{-\infty}^\infty \left(- \frac{m_{12}}{m_{22}} \right) \, da \exp \left( -\frac{m_{22}}{2} \left(a + \frac{m_{12}}{m_{22}} s\right)^2 - \frac{s^2}{2\sigma^2_s} \left( \frac{m_{11}m_{22} - m_{12}^2}{m_{22}} \right) \right) \nonumber \\
    &=  -\frac{m_{12}}{\sqrt{2\pi m_{22}^3 {\rm det} \bf \Sigma} }
 \int_\lambda^\infty s\, ds \exp \left( - \frac{s^2}{2\sigma^2_s} \right) \nonumber  \\
    &=  \frac{m_{12}}{m_{22}}\frac{1}{\sqrt{2\pi \sigma^2_s}}
 \sigma^2_s \exp \left( - \frac{\lambda^2}{2\sigma^2_s} \right) \nonumber \\
    &= \frac{{\rm Cov}(s, a)}{\sqrt{2\pi \sigma^2_s}}
 \exp \left( - \frac{\lambda^2}{2\sigma^2_s} \right) \label{eq:bias_steps}.
\end{align}

Finally, I use this result to compute the bias near a transition between two different weighting regimes.  Denoting the covariance above the threshold as ${\rm Cov}_+(s, a)$ and the covariance below the threshold as ${\rm Cov}_-(s, a)$, and assuming the difference between $s$ and its threshold value to be $\mu$, I have 
\begin{equation}
    \langle a - \tilde{a} \rangle = \left( \frac{{\rm Cov}_+ \left(s, a \right) - {\rm Cov}_- \left(s, a \right) }{\sqrt{2\pi \sigma^2_s}} \right) \exp \left( - \frac{\mu^2}{2\sigma^2_s} \right). \label{eq:bias_discrete}
\end{equation}
This is maximised when $s$ is at a transition between weighting schemes, in which case the exponent is zero.  

Two or more transitions between weighting regimes can contribute to the bias.  If two transitions contribute, the bias is given by a sum of integrals of the form
\begin{align}
    \langle a - \tilde{a} \rangle = \int_{-\infty}^{\mu} f_1(s)\,ds + \int_{\mu}^{\nu} f_2(s)\,ds + \int_{\nu}^{\infty} f_3(s)\,ds \label{eq:bias_multiplethresh_step1}
\end{align}
where, e.g., $f_1$ is the integrand in the first regime (c.f.~the first line of Equation \eqref{eq:bias_steps}), and $\mu$ and $\nu$ are the number of standard deviations the noiseless value of $s$ is away from a transition.  The functions $f_1$, $f_2$, etc.~are odd functions of $s$ (c.f.~the second line of Equation \eqref{eq:bias_steps}).  Using this fact, Equation \eqref{eq:bias_multiplethresh_step1} can be written
\begin{align}
    \langle a - \tilde{a} \rangle &= \int_{-\infty}^{\mu} f_1(s)\,ds + \int_{\mu}^{\infty} f_2(s)\,ds - \int_{\nu}^{\infty} f_2(s)\,ds + \int_{\nu}^{\infty} f_3(s)\,ds \nonumber \\
    &= \int_{\mu}^{\infty} \left( f_2(s) - f_1(s)\right) \,ds + \int_{\nu}^{\infty} \left( f_3(s) - f_2(s) \right)\,ds .
\end{align}
The first and second integrals each give the same result as Equation \eqref{eq:bias_discrete} across the two respective transitions.  The total bias is then the sum of Equation \eqref{eq:bias_discrete} over both transitions.  
This argument can be extended to show that the total bias is the sum of Equation \eqref{eq:bias_discrete} calculated over all transitions.

\begin{figure*}
\includegraphics[width=0.5\textwidth]{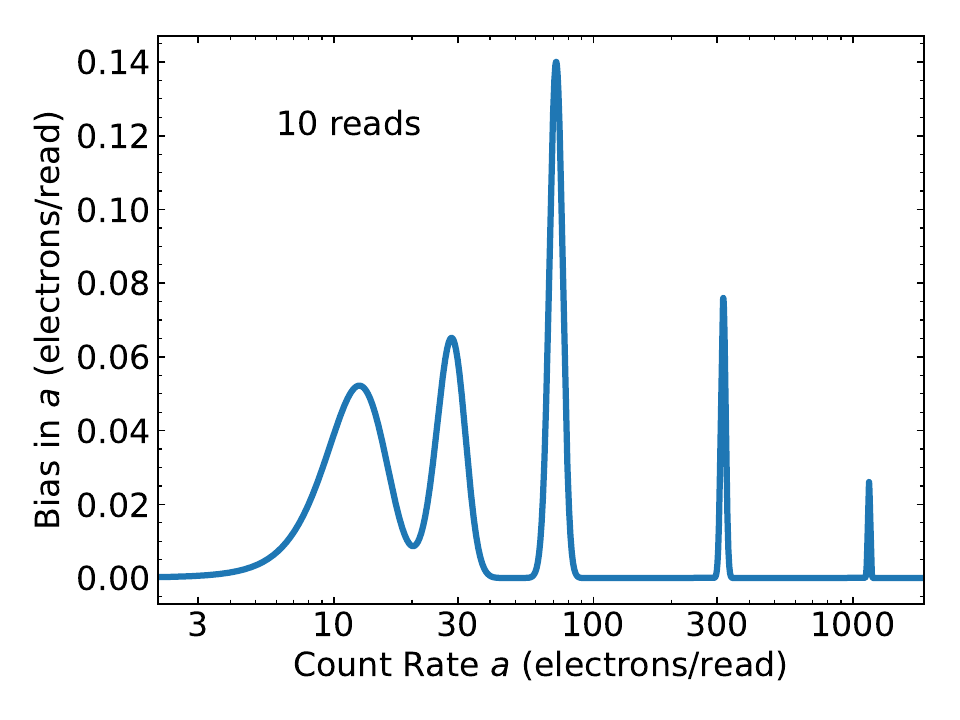}
\includegraphics[width=0.5\textwidth]{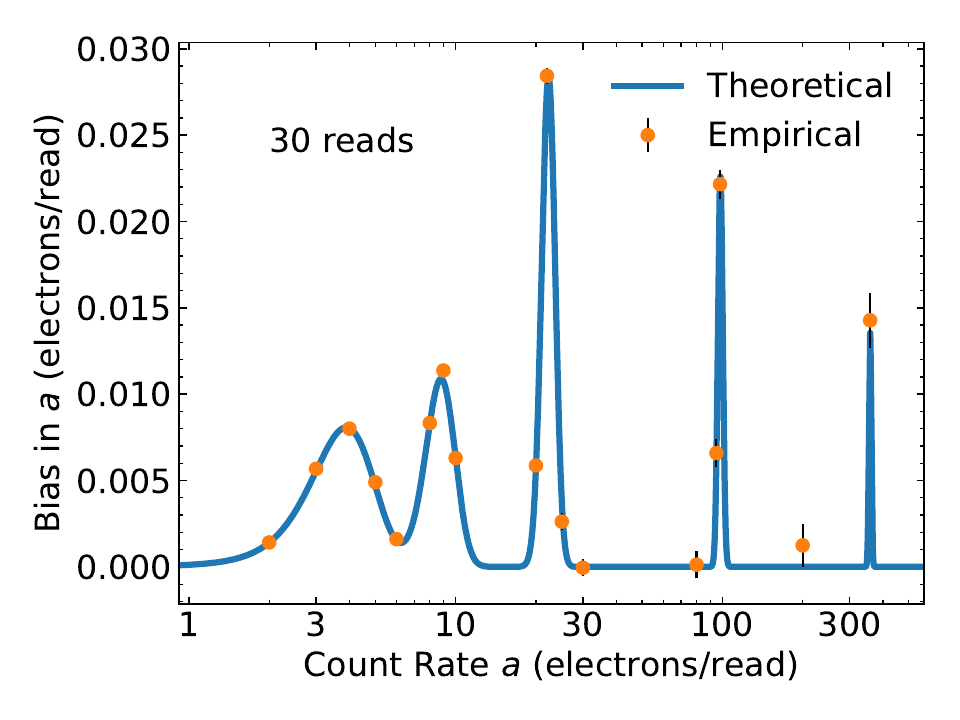}
\caption{The bias in the discrete weighting scheme of \cite{Fixsen+Offenberg+Hanisch+etal_2000} as adapted by \cite{Casertano_2022}, for 10 reads (left) and 30 reads (right) with a read noise of 20 electrons.  The orange points are the averages of $5 \times 10^6$ Monte Carlo realizations each with read and photon noise; they confirm the accuracy of the bias derived in this section. The bias is positive because a larger $s$, the difference between the first and last resultants, results in an increased weighting of these resultants in the computation of $a$, and a larger covariance between $s$ and $a$.  The peaks are centered at the transitions between weighting schemes while the widths of the peaks are given by $\sigma_s \approx \sqrt{s + \sigma^2}$ scaled to the total number of reads. 
 The width is a weakly increasing function of the count rate and appears to narrow because of the logarithmic axis. 
 \label{fig:bias_fixsen}}
\end{figure*}

Figure \ref{fig:bias_fixsen} shows the bias for ramps of 10 and of 30 reads.  For this calculation I have used the weighting schemes and signal-to-noise ratio thresholds given in \cite{Casertano_2022}, and have adopted a read noise of 20 electrons.  The bias is mostly due to the read noise component of the covariance.  For ten reads, the bias can be $\approx$0.5\% at count rates near the boundaries between different weighting schemes.  The orange points in Figure \ref{fig:bias_fixsen} show empirical calculations of the bias using Monte Carlo; they verify the accuracy of the theoretical curve derived in this section.  

As for the continuous case discussed below, the bias can be mostly removed from the discrete weighting schemes.  The difference in covariance between $a$ and $s$ is a linear combination of the weights (c.f.~Equation \eqref{eq:cov_a_s}).  The quantity $s$, the difference between the first and last reads, results in a biased estimate of $a$.  A general estimator for $a$ would have a covariance with $a$ that is likewise given by a linear combination of the weights $\{w_i\}$.  The dimensionality of this set of estimators is one less than the number of resultants.  Given a number of transitions $N$ less than the number of resultants, it is possible to choose an estimator for $a$ for which the covariance difference with $a$ is zero across all $N$ transitions.  Such an estimator would produce a nearly unbiased way of choosing which weights to apply.

\subsection{The Continuous Case}

I now derive the bias for a general least-squares fit as described in Section \ref{sec:fitramp} and show how to remove it.  I will start with Equation \eqref{eq:ramp_equation}, the formula for a ramp, and define
\begin{equation}
    w_i = \left(\sum_j \left({\bf C^{-1}}\right)_{ij}\right) \left(\sum_j \sum_k \left({\bf C^{-1}}\right)_{jk}\right)^{-1}
    \label{eq:weights_def}
\end{equation}
where the covariance matrix only applies to the resultant differences. 
 I then have 
\begin{align}
    a = \sum_i w_i d_i = {\bf w} \cdot {\bf d} \label{eq:weighted_sum_di}
\end{align}
and
\begin{align}
    \sum_i w_i = 1 . \label{eq:sum_wi}
\end{align}
The $w_i$ are themselves functions of the count rate $a$ assumed in the construction of the covariance matrix (for the photon noise portion).  For the rest of this discussion I will assume that the (unknown) actual count rate is $\tilde{a}$ while the covariance matrix is derived using a slightly different $a'$.  I will assume that the read noise associated with each pixel is accurately known.

I will first treat the case where the $a'$ used in the construction of the covariance matrix is not directly derived from any of the resultant differences $d_i$.  In this case, I can use the fact that 
\begin{equation}
    \langle d_i \rangle = \tilde{a} 
    \label{eq:eachdiff_unbiased}
\end{equation}
for all reads $i$ because the observed count rate is an unbiased estimator of the true count rate and because read noise has zero mean.  Equations \eqref{eq:weighted_sum_di}, \eqref{eq:sum_wi}, and \eqref{eq:eachdiff_unbiased} then imply that the $\chi^2$-minimizing fit gives an unbiased estimator of the flux:
\begin{align}
    \langle a \rangle &= \bigg \langle \sum_i w_i d_i \bigg \rangle \nonumber \\
    &= \sum_i w_i \langle 
 d_i \rangle \nonumber \\
 &= \sum_i w_i \tilde{a} \nonumber \\
 &= \tilde{a}.
 \label{eq:unbiased_default}
\end{align}
This does not hold if the $w_i$ depend on the values of $d_i$, i.e., if the $d_i$ values are used in determining the count rate for the purposes of deriving the covariance matrix.  In that case, I will assume that the covariance matrix is calculated assuming a photon rate of 
\begin{align}
    a' = \sum_i c_i d_i = {\bf c} \cdot {\bf d}
\end{align}
with 
\begin{align}
    \sum_i c_i = 1.
\end{align}
This is fairly general: if all $c_i$ are equal then this is the case of using the average count rate (scaled differences between adjacent groups of reads) to compute the covariance matrix; it is equivalent to using the difference between the first and last groups of reads.  Iteratively updating $w_i$ and estimating $a$ would correspond to another set of $c_i$ (different at each iteration).  Using weights for each resultant derived from the read-noise limited fit would correspond to a different set of $c_i$. 

The dependence of the weights $w_i$ on the adopted value of $a$ is complicated so I will use a Taylor expansion of $w_i$ to first order about the true count rate $\tilde{a}$.  I have
\begin{align}
    \langle a \rangle &\approx \Bigg\langle \sum_i \left( w_i(\tilde{a}) + \frac{dw_i}{da} \left( \left(\sum_j c_j d_j \right) - \tilde{a}\right)\right) d_i \Bigg\rangle \nonumber \\
     &= \tilde{a} + \Bigg\langle \sum_i d_i \frac{dw_i}{da} \left(\sum_j c_j \left( d_j - \tilde{a} \right) \right) \Bigg\rangle 
\end{align}
where I used $\sum c_j = 1$, $\langle d_i \rangle = \tilde{a}$, and $\sum w_i = 1$. 
I will further expand this by subtracting and adding $\tilde{a}$ to $d_i$: 
\begin{align}
    \langle a \rangle &\approx \tilde{a} + \Bigg\langle \sum_i \left( d_i - \tilde{a}\right) \frac{dw_i}{da} \left(\sum_j c_j \left( d_j - \tilde{a} \right) \right) \Bigg\rangle 
    + \tilde{a} \Bigg\langle \sum_i \frac{dw_i}{da} \left(\sum_j c_j \left( d_j - \tilde{a} \right) \right) \Bigg\rangle \nonumber \\
    &= \tilde{a} + \Bigg\langle \sum_i \sum_j c_j \frac{dw_i}{da} \left( d_j - \tilde{a} \right) \left( d_i - \tilde{a}\right) \Bigg\rangle 
    + \tilde{a} \Bigg\langle \sum_i \sum_j c_j \frac{dw_i}{da} \left( d_j - \tilde{a} \right) \Bigg\rangle \nonumber \\
    &= \tilde{a} + \sum_i \sum_j c_j \frac{dw_i}{da} \big\langle 
 \left( d_j - \tilde{a} \right) \left( d_i - \tilde{a}\right) \big\rangle 
    + \tilde{a} \sum_i \sum_j c_j \frac{dw_i}{da}  \big\langle \left( d_j - \tilde{a} \right) \big\rangle .
\end{align}
The last term is zero because the individual scaled resultant differences are unbiased estimators of the true count rate.  The first term has the covariance matrix of the resultant differences, ${\rm Cov}(d_i, d_j)$, given by ${\bf C}$ in Equation \eqref{eq:covar_sum}:
\begin{align}
    \langle a \rangle &= \tilde{a} + \sum_i \sum_j c_j \frac{dw_i}{da} {\rm Cov}(d_i, d_j) . \label{eq:bias}
\end{align}
So, if we adopt a covariance matrix built using a weighted sum of the resultant differences to estimate the photon rate, then Equation \eqref{eq:bias} gives a first-order estimate of the bias introduced to the recovered count rate.  It is possible to either correct for this bias or to choose a set of weights $c_j$ for the initial estimate of the count rate in order to have zero bias to first order.  If we want to avoid the bias, then we wish to choose a vector of initial guess coefficients ${\bf c}$ so that ${\bf c}$ is orthogonal to
\begin{equation}
    {\bf v} \equiv {\bf C} \frac{d{\bf w}}{da} . \label{eq:bias_vec}
\end{equation}

In fact, the set of optimal coefficients ${\bf w}$ to combine the resultant differences is a bias-free choice for ${\bf c}$.  To prove this I will use the fact that the weights ${\bf w}$ given in Equation \eqref{eq:weights_def} provide the minimum-variance unbiased estimate of the true count rate if the true covariance matrix is ${\bf C}$ \citep{Aitken_1935}.  The variance of the sum of resultant differences weighted by ${\bf w}$ is the variance of the recovered count rate $a$, and is given by
\begin{equation}
    \sigma^2_a = {\bf w}^T {\bf C} {\bf w}.
\end{equation}
A weight vector ${\bf w}'$ derived with a different assumed count rate (i.e.~a different approximation to the true covariance matrix) will still produce an unbiased estimate of the count rate due to Equation \eqref{eq:unbiased_default}.  In other words, ${\bf w}(a)$ gives an unbiased estimate of the true count rate for any assumed count rate $a$ used to approximate the covariance matrix and, from this, compute ${\bf w}$ using Equation \eqref{eq:weights_def}.  The Gauss-Markov theorem then states that $\sigma^2_a$ is minimized if the weight vector ${\bf w}$ is derived using the true count rate $a_{\rm true}$.  So, differentiating $\sigma^2_a$ with respect to the count rate used to derive ${\bf w}$ will equal zero at $a_{\rm true}$:
\begin{align}
0 &= \frac{d}{da} \left({\bf w}^T {\bf C} {\bf w}\right) \nonumber \\
&= \left(\frac{d{\bf w}}{da} \right)^T {\bf C} {\bf w} + {\bf w}^T {\bf C} \frac{d{\bf w}}{da} \nonumber \\
&= 2 {\bf w}^T {\bf C} \frac{d{\bf w}}{da}
\label{eq:weights_unbiased}
\end{align}
where the last line used the symmetry of the covariance matrix ${\bf C} = {\bf C}^T$.  So, if the optimal weight vector ${\bf w}$ can be approximately calculated, then the covariance matrix computed from $a = {\bf w} \cdot {\bf d}$ allows for a nearly unbiased estimate of the true count rate.  If the covariance matrix is approximated using $a = {\bf c} \cdot {\bf d}$ for some other ${\bf c}$, then the resulting best-fit count rate will be biased by an amount
\begin{equation}
    {\rm bias} \approx {\bf c}^T {\bf C} \frac{d{\bf w}}{da} .
    \label{eq:bias_formula_init}
\end{equation}

The bias of Equation \eqref{eq:bias_formula} results from a series expansion of ${\bf w}$ about the true count rate $a$.  Negative values of $a$ are incompatible with the Poisson distribution; the covariance matrix should not have a negative coefficient times the photon noise covariance matrix.  In practice this means that Equation \eqref{eq:bias_formula_init} overestimates the bias when the count rate is close to zero assuming that the covariance matrix is approximated using the maximum of $a={\bf c} \cdot {\bf d}$ and zero.  We can estimate this effect using the probability that the initial weight vector will produce a negative estimated count rate, and reduce the bias by this factor.  Assuming Gaussian errors and an initial uncertainty on the count rate of
\begin{equation}
    \sigma^2_0 = {\bf c}^T {\bf C} {\bf c},
\end{equation}
we finally have
\begin{equation}
{\rm bias} \approx \frac{1}{2} \left(1 + {\rm erf} \left( \frac{a}{\sigma_0 \sqrt{2}} \right) \right) {\bf c}^T {\bf C} \frac{d{\bf w}}{da} \label{eq:bias_formula}
\end{equation}
where the initial factor is the probability of obtaining a negative measured value assuming a true value of $a$ and an uncertainty of $\sigma_0$.

\subsection{Empirical Demonstrations of the Bias} \label{sec:bias_empirical}

I have tested the first-order approximation for the bias on synthetic data with 30 reads each treated individually.  The off-diagonal elements of the covariance matrix in this case consist only of read noise.  I further adopt a read noise of $\sigma=20$\,$e^-$/read.  The bias will depend upon the actual count rate (which partially determines the covariance matrix ${\bf C}$) and on the weight vector ${\bf c}$ used to estimate the count rate for use in approximating the covariance matrix.  

\begin{figure}
    \centering \includegraphics[width=0.9\linewidth]{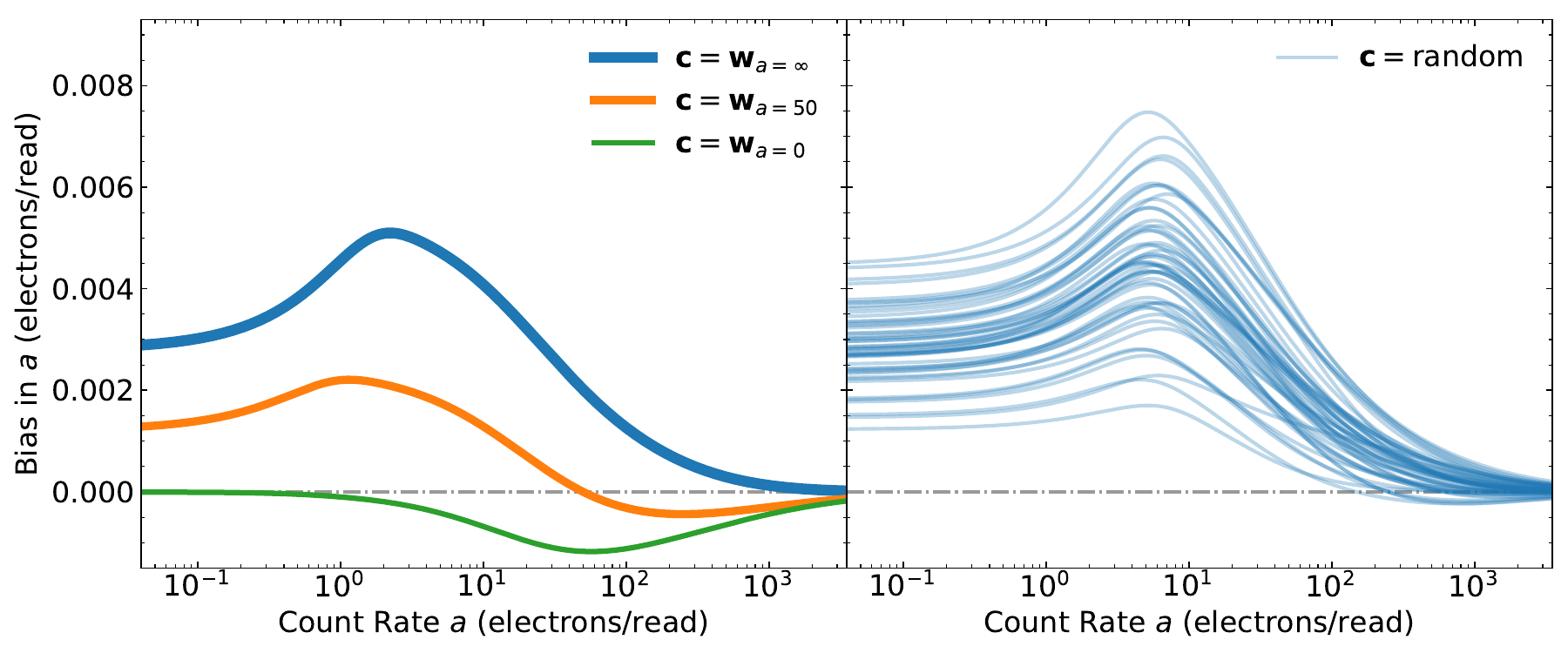}
    \caption{Biases computed using Equation \eqref{eq:bias_formula} for a sequence of 30 individual reads with a read noise of 20 electrons using various initial weight vectors ${\bf c}$.  Left panel: the ${\bf c}$ vectors are set to the optimal weights for three different true count rates.  Right panel: biases for 50 random vectors ${\bf c}$ with all elements drawn from $U(0, 1)$ and the vector finally normalized to a unit sum.  The bias can be significant at low count rates depending on the weights on the resultant differences used to estimate the covariance matrix.  The bias is zero when the vector ${\bf c}$ is the optimal weight vector for the actual count rate. }
    \label{fig:biases}
\end{figure}

Figure \ref{fig:biases} plots the bias computed using Equation \eqref{eq:bias_formula} as a function of count rate for several different initial weight vectors ${\bf c}$.  The left panel shows the bias resulting from the optimal weight vector for zero count rate, a moderate count rate of 50 electrons/read, and an arbitrarily high count rate for which all elements of ${\bf c}$ are the same.  The right panel of Figure \ref{fig:biases} shows the bias resulting from 50 random realizations of the initial weight vector ${\bf c}$.  In all cases, I use uniform random numbers between zero and one for all elements and then normalize the vector so that the elements sum to one.  The biases in both cases can be non-negligible at low count rates.  

\begin{figure*}
    \centering\includegraphics[width=0.5\textwidth]{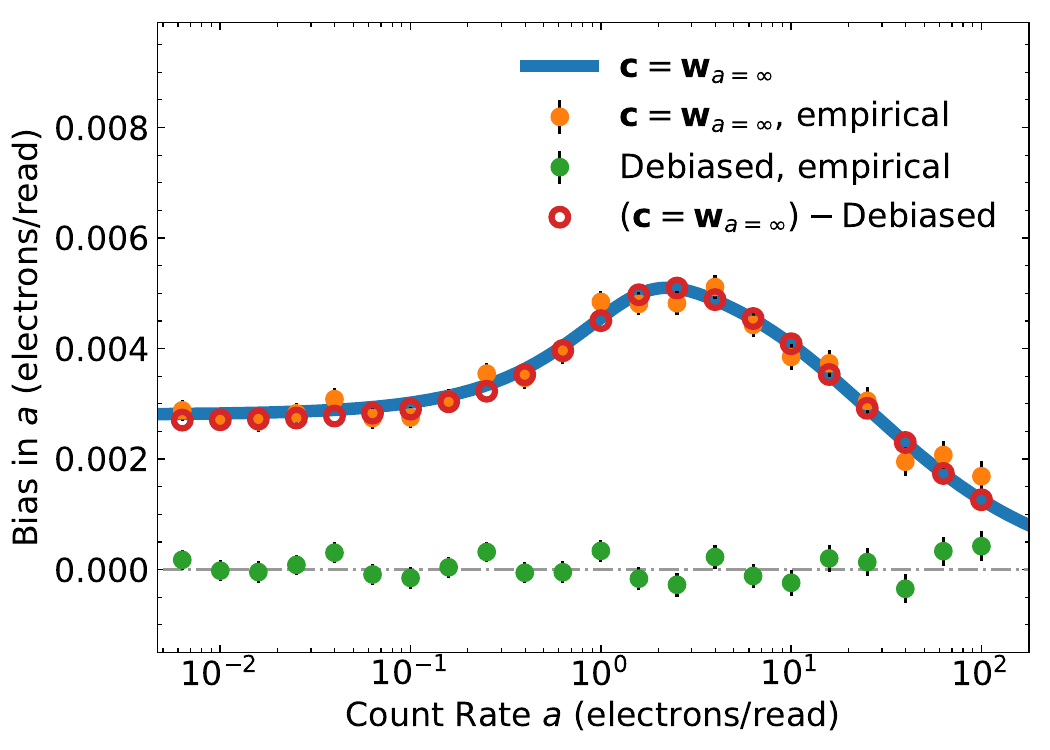}
    \caption{Verification of the bias computed using Equation \eqref{eq:bias_formula} (blue line) using Monte Carlo (orange points) assuming 30 reads with 20 electron read noise (as in Figure \ref{fig:biases}).  The green points use the same ramps as the orange points, but fit the ramp twice as suggested in Section \ref{sec:bias_empirical}. The bias after fitting the ramp twice is negligible. The open red points show the difference between the orange and green points, i.e., the empirical bias assuming that fitting the ramp twice produces an exactly unbiased fit.  Uncertainties on the open red points are negligible.
\label{fig:bias_empirical}}
\end{figure*}

Next, I test the bias calculated using Equation \eqref{eq:bias_formula} with Monte Carlo.  For this I continue to assume 30 individual reads with a read noise of 20 electrons.  Figure \ref{fig:bias_empirical} compares the bias from an initial estimate using only the first and last reads (i.e.~averaging the read differences) to fits of Monte Carlo realizations of ramps.  Using only the first and last reads provides the optimal estimate at high photon rates, and we therefore expect it to produce unbiased count rates in this regime.  At low photon rates, however, this weighting is not optimal and Equation \eqref{eq:bias_formula} predicts a bias.  I generate at least $5 \times 10^6$ synthetic ramps at each sample count rate to provide a robust empirical bias; this is indicated by the orange points in the left panel.  The Monte Carlo results agree well with the prediction for count rates $\gtrsim$1 electron/read.  The points also agree well at low count rates where the correction factor in Equation \eqref{eq:bias_formula} approaches $\frac{1}{2}$.

I can remove the bias to first order by computing Equation \eqref{eq:bias_vec}, projecting this vector off of my initial weight vector, renormalizing the weight vector, and repeating the Monte Carlo test.  This does not give the optimal weight vector but it does give one that will produce an estimate of the covariance matrix that results in unbiased fitted count rates.  Using this approach with $10^7$ synthetic ramps and a true count rate of 2, the mean best-fit slope becomes 2.00015 with an uncertainty on the mean of 0.00016, i.e., the bias is more than 10 times lower and is no longer detectable without running a much larger set of synthetic ramps.  

I can also avoid almost all of the bias by performing the fit to the ramp twice.  I use the first fit to infer the weights ${\bf w}$, and after using these weights to estimate the photon rate, I recompute the covariance matrix.  I then use this new covariance matrix to perform a second fit to the ramp in order to compute the final count rate.  

The green points in the left panel of Figure \ref{fig:bias_empirical} are fits to the same ramps as the orange points, but fit the ramp twice to remove bias.  The red points in the right panel of Figure \ref{fig:bias_empirical} show the difference between the orange and green points in the left panel.  They show the bias from using only the first and last reads to estimate the covariance matrix, {\it assuming} that fitting the ramp twice produces unbiased results.  These red points agree almost perfectly with the prediction at high count rates, and agree almost as well at low count rates where the correction factor for nonnegative inferred count rates becomes significant.

To obtain the best estimate of the true photon rate and avoid biases in the process, I therefore suggest the following procedure:
\begin{enumerate}
    \item Use uniform weights or a median on all scaled resultant differences $d_i$ to estimate a count rate;
    \item Use this count rate to estimate the covariance matrix and fit for the count rate; 
    \item Use this updated count rate to re-estimate the covariance matrix; and
    \item Perform the optimal fit with this re-estimated covariance matrix.
\end{enumerate}
The total computational cost of this approach is approximately double the cost of fitting the ramp once.

\section{Example: a NIRCam RAPID Exposure}

I demonstrate the new ramp fit on NIRCam RAPID data with eight reads, and a single read per group.  These data are from Early Release Science (ERS) imaging of NGC 3324 in the F200W filter with the {\tt nrca1} detector.  The total exposure time was 161 seconds.  A visual comparison of the ramps themselves (i.e.~images of count rates at each pixel) requires comparable jump detection and masking; I defer this comparison to \citetalias{Brandt_2024} and do not show any images here.  

In this paper I address the bias and $\chi^2$ statistics of the ramp fit as visible in the count rates available in the {\tt \_rate.fits} file available on MAST.  I perform no bias subtraction or nonlinearity correction to the groups.  I do, however, correct for the reference pixels.  I subtract the average value of the reference pixels at either end of each of the four readout channels and then use the reference pixels along the sides of the detector in Channels 1 and 4 to remove some of the $1/f$ noise.  I smooth these reference pixels with a Gaussian and subtract the pattern from the each channel, choosing the smoothing length and the factor by which I subtract to minimize the scatter about the resulting ramp fit.  I then adopt the read noise files available on the JWST calibration center, dividing by $\sqrt{2}$ to convert from correlated double sampling (CDS) noise to single read noise, and I use the calibration gain of 2.05\,$e^-/{\rm DN}$.

\begin{figure}
    \centering\includegraphics[width=0.5\textwidth]{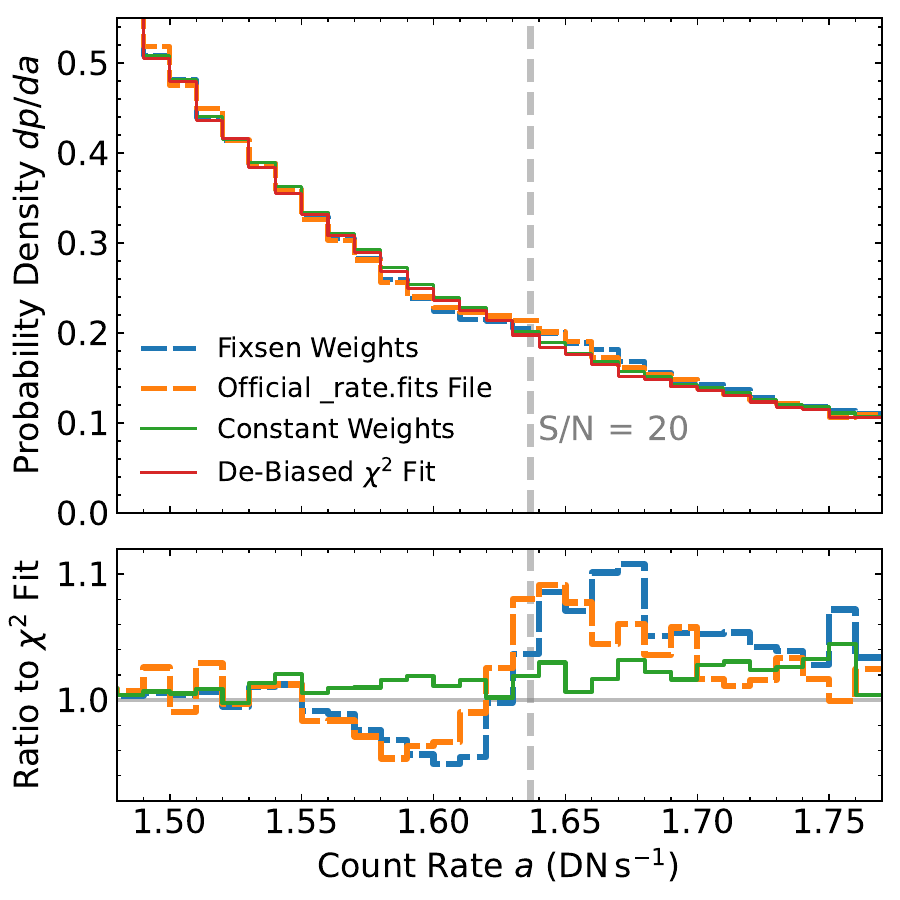}
    \caption{Demonstration of nonzero bias in actual JWST data from NIRCam.  The top panel shows the probability density of count rates $a$ using different ramp fitting approaches.  The bottom panel normalizes these probability densities to the ones I obtain with the approach described in this paper (red line in the top panel).  The MAST {\tt \_rate.fits} file uses the \cite{Fixsen+Offenberg+Hanisch+etal_2000} discrete weights.  The bias is visible in both a custom ramp fit using the \cite{Fixsen+Offenberg+Hanisch+etal_2000} weights (blue dashed line) and in the {\tt \_ramp.fits} file (orange dashed line) at the expected location, where the S/N crosses the threshold between weighting schemes.  Each pixel has a different noise level, a fact that blurs the bias over a broader range of count rates.  The bias is not visible when using weights that do not change at this threshold (green line) or when using a debiased fit (red line).  \label{fig:empirical_bias}}
\end{figure}

Section \ref{appendix:bias} suggests that the {\tt \_rate.fits} file derived from the individual groups may have a detectable bias from the discrete change in the weights.  Figure \ref{fig:empirical_bias} shows that this is indeed the case.  The figure plots a histogram of the count rates near a signal-to-noise ratio (S/N) of 20 for a pixel with a typical noise level; the weighting scheme changes at this S/N.  The {\tt \_rate.fits} file has a small hump just past this value, which my new debiased ramp file lacks.  A new ramp fit with the original weights given in \cite{Fixsen+Offenberg+Hanisch+etal_2000} confirms this as the source of the bias.  An alternative weighting scheme that does not change near this S/N also does not show a bias (green histogram).  The lower panel shows the ratio of the histograms of the biased ramps and of the constant weight ramp to that of the unbiased $\chi^2$ ramp.  The biased ramps show a deficit of points just below the threshold and an excess just above due to the shifting of individual ramp fits to slightly higher values.  

\begin{figure}
    \includegraphics[width=0.5\textwidth]{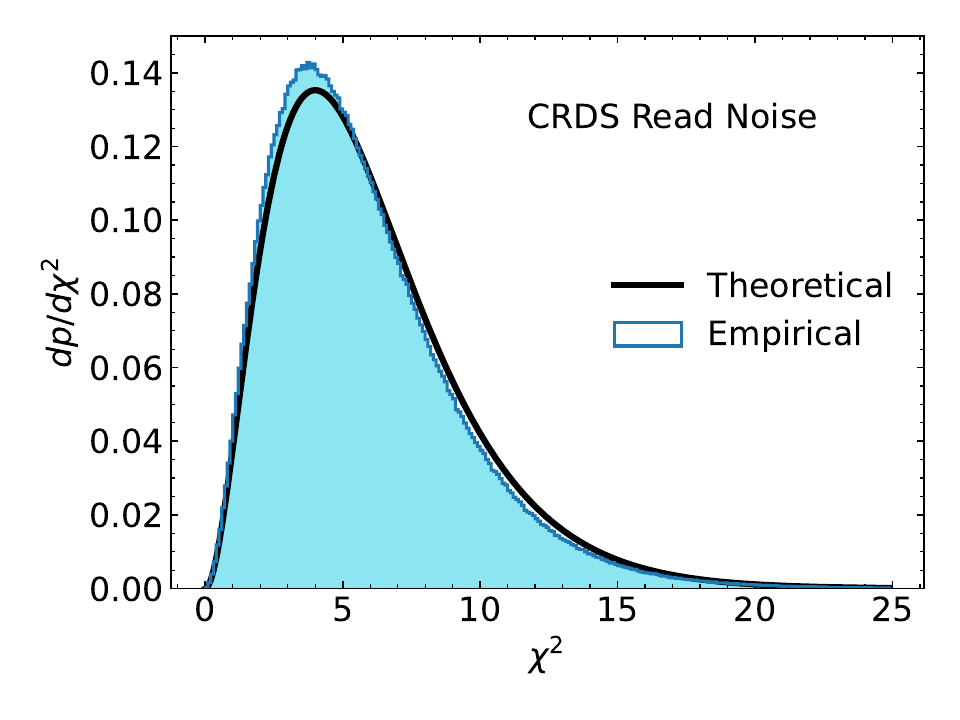}
    \includegraphics[width=0.5\textwidth]{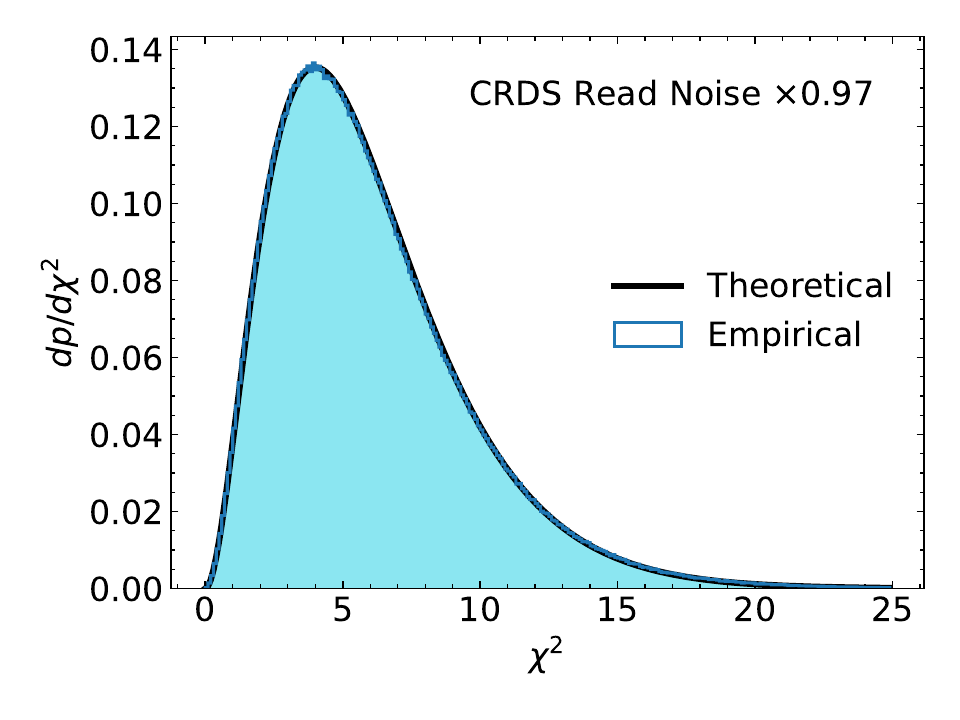}
    \caption{Distribution of $\chi^2$ values for a NIRCam image taken in RAPID mode.  The left panel uses read noise levels from the Calibration Reference Data System; in the right panel the noise values have been scaled by a factor of 0.97 (i.e.~slightly reduced) to obtain the best agreement with a theoretical distribution.  The theoretical distribution is a $\chi^2$ distribution with six degrees of freedom (seven group differences minus one fitted parameter).  The theoretical and empirical distributions are indistinguishable when scaling the read noise. The $\approx$4\% of pixels with a detected jump, identified as described in \citetalias{Brandt_2024}, are excluded.  
    \label{fig:chisq}}
\end{figure}

Finally, the algorithms described in this paper provide a direct measurement of the $\chi^2$ value for the fit.  Figure \ref{fig:chisq} shows a histogram of the $\chi^2$ values for the ramp fit compared to a theoretical distribution with six degrees of freedom: seven group differences minus one fitted slope.  The left panel adopts the read noise values from the JWST Calibration Reference Data System\footnote{\url{https://jwst-crds.stsci.edu/}}.  In this right panel I have scaled the noise down slightly, by a factor of 0.97, for better agreement.  The noise scaling depends on how well the reference pixel correction removes correlated noise, leading to slight differences between my favored noise values and those in the JWST calibration package.  This result suggests that some improvement in the reference pixel correction used by JWST may be possible.  With this slight scaling of the noise, the empirical distribution of $\chi^2$ values is indistinguishable from the theoretically expected distribution.  This suggests that the adopted covariance matrix provides a very good statistical description of the data, and validates the formal uncertainties of the fitted slopes.

\section{A Pure Python Implementation} \label{sec:implementation}

I have implemented the algorithms described in this paper in pure Python.  In this section I briefly summarize the implementation and its computational cost.  All tests running the code were performed on a 2020 Macbook Air.  I have further included a series of tests to verify that all calculations are correct: that the calculated covariance matrix agrees with a Monte Carlo approximation and that the best-fit slopes agree with the results of explicit matrix inversion.

The first step in my implementation is to compute the $\alpha$ and $\beta$ components of the covariance matrix and the $\delta t$ values for a set of read times.  This set of read times is a list of the time(s) since reset, or the integration time(s), for the read(s) corresponding to each resultant.  A single read resultant may be specified by either a floating point number for the integration time or a list of numbers for a multiple-read resultant.  The resulting $\alpha$ and $\beta$ components for photon noise and read noise are then stored in a specifically designed Python class.  A calling sequence for a six resultant ramp with a mixture of single-read resultants and multiple-read resultants could look like the following:
\needspace{3\baselineskip}
\begin{lstlisting}[language=Python,numbers=none]
import fitramp
readtimes = [1, 2, [3, 4, 5], [6, 7, 8], [10, 11, 13], 15]
C = fitramp.Covar(readtimes)
\end{lstlisting}
If the user would like to fit for and/or apply an informative prior on the pedestal value, they would call {\tt fitramp.Covar} with the boolean {\tt pedestal} set to {\tt True}.  The covariance structure would then have an extra element in both $\alpha$ and $\beta$ as described in Section \ref{sec:fit_reset}. 

The next step is to fit a ramp.  The corresponding function takes a 2D array of resultant differences (number of resultants minus one by number of pixels), the Python class holding the covariance information from the integration time of each read, and the read noise of each pixel.  It is vectorized to operate on many pixels simultaneously.  Optionally, this step may include a mask of the same shape as the resultant differences (differences with a mask value of zero are ignored), and it can compute count rates and $\chi^2$ values leaving out resultant differences and pairs of differences.  These quantities may be used in a jump detection approach, described in \citetalias{Brandt_2024}.  If the pedestal value is to be used, the first element of the resultant difference array should be the first resultant divided by its integration time. 
Finally, I have included a Python method to compute the bias in the count rate from using a weighted average of the resultant differences to estimate the count rate for the covariance matrix.  

Implementing the ramp fitting algorithm described in this paper requires computing a number of auxiliary quantities.  These quantities, defined throughout this paper, all have a linear cost in the number of resultants.  They also require memory.  To limit the memory footprint, I recommend using the ramp fitting algorithm on one row of detector pixels at a time.  A row-by-row loop also enables more efficient memory access compared to trying to access large parts of many different arrays that exist in distant regions of RAM.  In practice I have found maximal efficiency from operating on $10^3$--$10^4$ pixels at a time.  This approach leaves a negligible memory footprint.  

\begin{figure}
    \centering\includegraphics[width=0.5\textwidth]{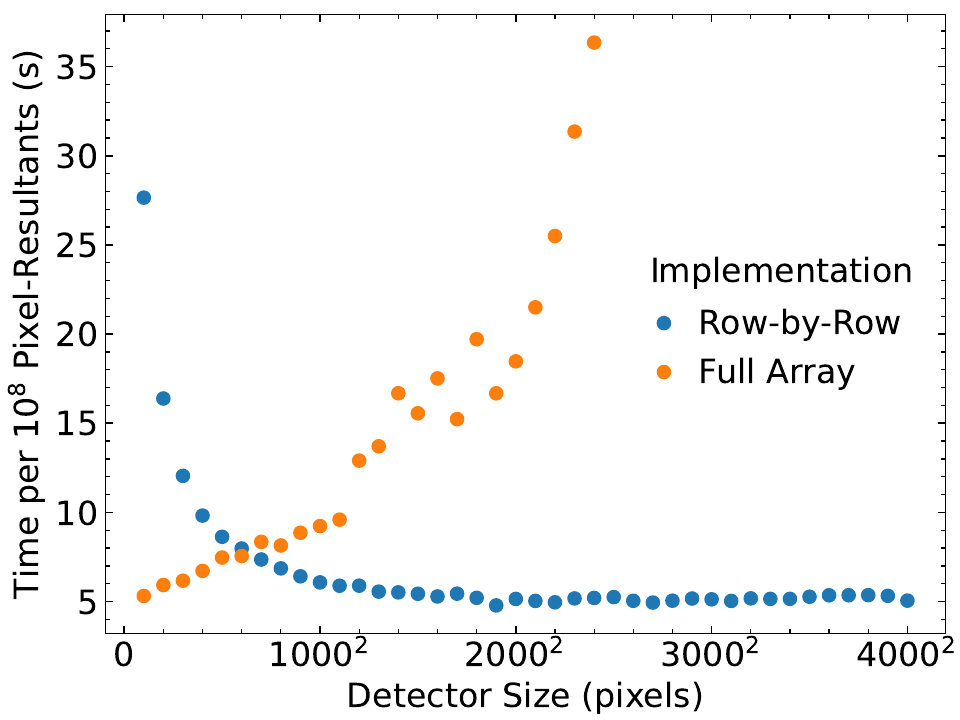}
    \caption{Computational time per pixel and per resultant of an up-the-ramp fit with ten resultants, fitting twice to remove bias, with a row-by-row implementation (blue points) compared to a single pass on the full array (orange points).  The single pass is faster when the detector is small, but as the detector grows larger, arrays for all of the auxiliary quantities begin to demand all of the system's RAM and performance suffers.  This crossover point will be machine dependent.  The row-by-row implementation is better for large-format detectors processed using a laptop or desktop computer (the tests were run on a computer with 8 GB RAM). \label{fig:runtime}}
\end{figure}

Figure \ref{fig:runtime} shows the computational time for ten resultant ramps using both the row-by-row implementation and by operating on the full array at once.  The full array approach becomes costly as the memory required for auxiliary quantities approaches the system's RAM; this point will be machine dependent but happens for an array size of $\sim$
2000$^2$ pixels on my laptop (with 8 GB of RAM).  A row-by-row implementation remains efficient for 4000$^2$-pixel detectors and has a negligible memory footprint beyond that required to store the resultants themselves.

The computational cost of this approach, while larger than that of \cite{Fixsen+Offenberg+Hanisch+etal_2000} and especially its memory-efficient implementation by \cite{Offenberg+Fixsen+Mather_2005}, is modest for a modern computer and is linear in the number of resultants.  For that reason the performance numbers I quote, and those shown in Figure \ref{fig:runtime}, are in units of seconds per $10^8$ pixel-resultants.  A ramp with twice as many resultants will take twice as long to process, as will a ramp with twice as many pixels.  A ramp with $10^8$ pixel-resultants roughly corresponds, for example, to an H2RG ($\approx 4 \times 10^6$ pixels) with 24 resultants.

Running my pure Python implementation on a single core of a 2020 Macbook Air takes $\approx$2.6 seconds per $10^8$ pixel-resultants to fit a ramp once.  Fitting a ramp twice to remove bias doubles this cost to a little over five seconds per $10^8$ pixel-resultants.  For an H4RG ramp with 10 resultants this cost corresponds to $\sim$8 seconds to fit a ramp and remove bias. These times correspond to a 3-year-old laptop and could be considerably lower on a better computer.  They would be correspondingly lower for ramps taken from a smaller H2RG detector.

\subsection{Numerical Considerations}

The computations needed to calculate the best-fit slope, its uncertainty, and $\chi^2$ are given by the equations of Section \ref{sec:fitramp}.  For a diagonally dominant matrix (as all covariance matrices that will arise for realistic ramps are) these typically do not present numerical difficulties.

If the photon rate and/or read noise are large, then overflow is a risk.  For example, assuming one read per second,
\begin{equation}
    \theta_n \sim \left( {\rm max}\left(a, \sigma^2\right) \right)^n .
\end{equation}
If $n=100$ and $\sigma^2 = 10000$ (for a very long ramp with a very noisy pixel) then overflow could result.  Overflow could occur for similar reasons in the recursive computation of auxiliary quantities.  By default, my implementation factors the geometric mean of $\alpha$ out of the covariance matrix to guard against overflow or (less likely) underflow.  This does not affect the best-fit slope.  After computation, the uncertainty on the best-fit slope is multiplied by the square root of this scaling factor and the value of $\chi^2$ is divided by the scaling factor.  

\section{Conclusions} \label{sec:conclusions}

Past work in the literature has either approximated the optimal solution to the problem of fitting a ramp \citep{Fixsen+Offenberg+Hanisch+etal_2000,Kubik+Barbier+Chabanat+etal_2016,Casertano_2022}, or has required expensive matrix operations \citep[e.g.][]{Robberto_2014}.  Here I have shown that the optimal approach can be implemented with a computationally efficient algorithm.  Closed-form solutions for the weights of the resultants are available, and the computational costs are linear in the number of resultants.  The optimal approach does require the covariance matrix of the resultants to be estimated first; this can introduce a bias in the best-fit count rate.  I have derived a formula for the bias and shown how it can be removed to first order.

As a byproduct of deriving the optimal count rates, I have also shown that the $\chi^2$ values of the fits may be computed for little additional cost.  This enables straightforward flags for the goodness of fit, which may be used to identify bad pixels.  The distribution of $\chi^2$ values may also be used to verify the quality of the noise model.

The algorithms presented here can be implemented efficiently in pure Python.  They are computationally straightforward on a laptop computer even for long ramps on a large-format detector.  This could enable more straightforward and sensitive ramp fitting for existing and future instruments using detectors that are read out nondestructively.

\software{scipy \citep{2020SciPy-NMeth},
          numpy \citep{numpy1, numpy2},
          Jupyter (\url{https://jupyter.org/}).
          }

\begin{acknowledgements}
I thank Stefano Casertano and Eddie Schlafly for helpful input and suggestions, and Sanjib Sharma, Michael Regan, and Karl Gordon for useful conversations. 
\end{acknowledgements}

\bibliography{uptheramp}{}
\bibliographystyle{aasjournal}

\appendix

\section{Nonuniform Weighting Within a Resultant}

\label{sec:intraresultantweights}

The analysis in Sections \ref{sec:covmatrix} and \ref{sec:fitramp} assumes that the individual reads within a resultant are averaged, with each read contributing equally.  This does not have to be the case, and a weighted average of the reads can offer better performance.  In this section I will focus on the optimal weighting in the read noise limited case.  These weights are straightforward to implement, always outperform uniform weighting, and remain compatible with suppression algorithms for correlated read noise and cosmic rays.  In the limit of low signal, this approach achieves the same signal-to-noise ratio on read noise limited data as saving all of the reads. 
The quantitative discussion of jumps in \citetalias{Brandt_2024} uses the case of equally weighted resultants, though the formulas in Section \ref{sec:covmatrix} straightforwardly generalize to nonuniform weights; the formulas and approach of Sections \ref{sec:fitramp}, \ref{sec:bias}, and of \citetalias{Brandt_2024} would be identical.

I will start by deriving the optimal coefficients for an up-the-ramp fit with $N$ reads at $N$ times $\{t_1, \ldots, t_N\}$ assuming only read noise.  The number of counts in a pixel, neglecting noise, should then be
\begin{equation}
    y_i = a t_i + b .
\end{equation}
I will write down $\chi^2$ as
\begin{align}
    \chi^2 = \sum_i \frac{\left(y_i - at_i - b\right)^2}{\sigma^2}.
\end{align}
I can find the best-fit slope by differentiating $\chi^2$ with respect to $a$ and $b$ and setting the derivatives equal to zero.  The solution for $a$ may be written
\begin{align}
    a = \frac{1}{N \left( \overline{t^2} - \overline{t}^2\right)} \left( \sum_i t_i y_i - \overline{t} \sum_i y_i \right) 
\end{align}
with
\begin{align}
    \overline{t} &\equiv \frac{1}{N} \sum_i t_i \\
    \overline{t^2} &\equiv \frac{1}{N} \sum_i t_i^2 .
\end{align}
The coefficient that multiplies each read is then
\begin{align}
    c_i = \frac{t_i - \overline{t} }{N \left( \overline{t^2} - \overline{t}^2\right)} .
\end{align}
If there is more than one read in a resultant, optimal intra-resultant weights may be defined by
\begin{align}
    \kappa_i &= \frac{c_i}{\sum_{i \in {\rm res}} c_i} \nonumber \\
    &= \frac{t_i - \overline{t}}{\sum_{i \in {\rm res}} \left(t_i - \overline{t}\right)} 
    \label{eq:proposedweights}
\end{align}
where $\overline{t}$ refers to the average read time of all of the reads in the ramp.  The total weight within a resultant is constrained to be one, and if there is only one read in a resultant, it will continue to have unit weight. 

Each resultant will now be a weighted average of reads.  The covariances derived in Section \ref{sec:covmatrix} must therefore be generalized.  For read noise the generalization is straightforward, with
\begin{equation}
    {\rm Var} ({\bf r}_i) =  \sigma^2 \sum_j \kappa_{i,j}^2,
\end{equation}
where $j$ runs over the reads within resultant $i$, taking the place of Equation \eqref{eq:readnoisevar}.  For photon noise, the covariance between resultant $i$ and resultant $i+1$ is 
\begin{equation}
    {\rm Cov}({\bf r}_i, {\bf r}_{i+1}) = a \left( \sum_{j} \kappa_{i,j} t_{i,j} \right) \equiv a \langle t_i \rangle
\end{equation}
which simplifies to Equation \eqref{eq:covar_rirj} if the $\kappa$ values are all equal (in which case the new and old definitions of $\langle t_i \rangle$ are equivalent).  The variance of resultant $i$ due to photon noise is 
\begin{equation}
    {\rm Var}({\bf r}_i) = a \sum_j \sum_k \kappa_{i,j} \kappa_{i,k} {\rm min}(t_{i,j}, t_{i,k}) .
\end{equation}
This expression takes the place of Equation \eqref{eq:var_singleres_phnoise}, which can no longer be simplified.  These equations may be propagated through the remainder of the derivations in Section \ref{sec:covmatrix} to derive the appropriate values for $\alpha$ and $\beta$ that define the covariance matrix.

The covariance matrix will remain tridiagonal unless a resultant has weights that sum to zero.  In that case, Equation \eqref{eq:proposedweights} is undefined.  Adopting Equation \eqref{eq:proposedweights} without the denominator would not solve the problem.  Two of the four terms in Equation \eqref{eq:zerocov_nonadjacent} (either the first two or the last two) would be zero and the sum would no longer vanish.  This numerical problem may be avoided by slightly changing the weight of one of the reads while keeping $\sum \kappa_i = 1$; a small change to the weights entails a negligible penalty in performance.

\begin{figure*}
    \includegraphics[width=\textwidth]{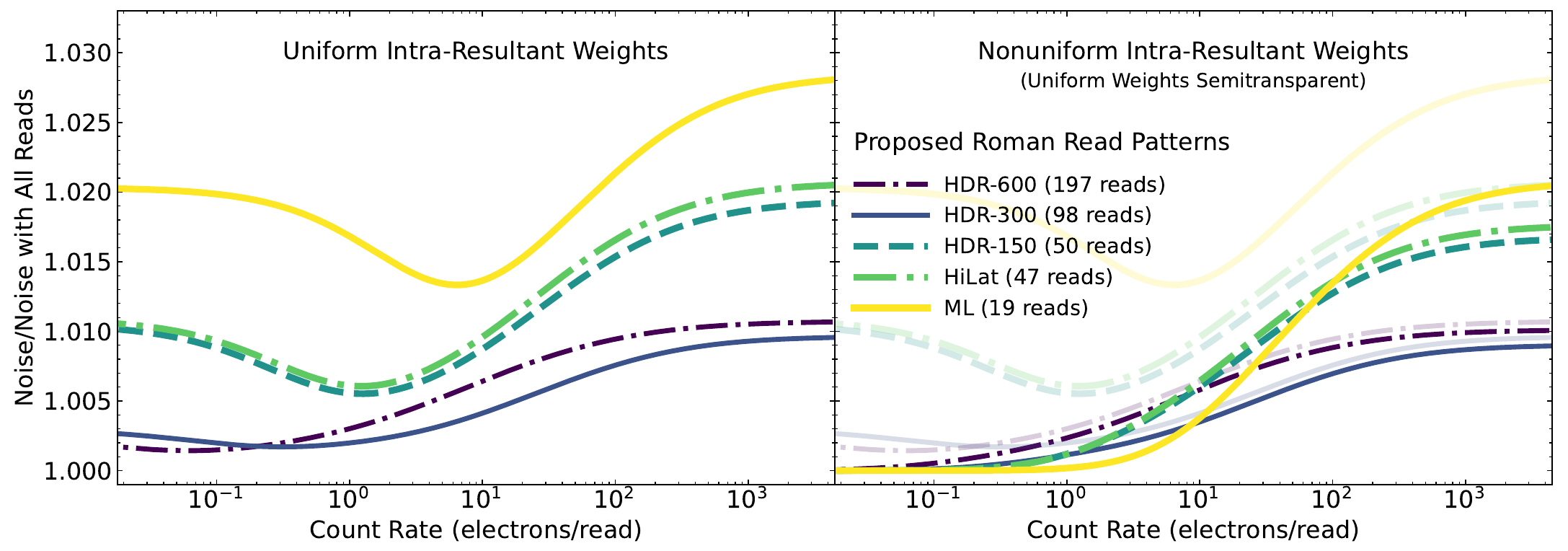}
    \caption{Left: Ratio of optimal slope fit to the optimal slope fit with all of the reads assuming uniform intra-resultant weights.  The different colors show the same proposed Roman readout patterns used in Figure \ref{fig:snr_compare}.  Right: the same ratio but with intra-resultant weights optimized in the low count rate regime.  The signal-to-noise ratio now reaches the value when using all of the reads for low count rates, and it is superior to the signal-to-noise for uniform intra-resultant weights at all count rates. The curves from the left panel are shown semitransparent in the right panel to facilitate a visual comparison. 
 \label{fig:performance}}
\end{figure*}

I can now measure the performance of uniform weights against the performance of nonuniform weights within each resultant.  Figure \ref{fig:performance} shows the results for the same proposed Roman readout patterns shown in Figure \ref{fig:snr_compare}.  Using the intra-resultant weights given in Equation \eqref{eq:proposedweights} matches the signal-to-noise ratio from using all of the reads in the low count rate limit.  At all count rates, it gives superior results to uniform intra-resultant weighting.  This is because in all cases the best weights increase toward the first and last reads; the optimal low count rate weights give the most gradual increase in weights toward either end.  This approach is still closer to the optimal photon noise limit of only using the first and last read than is the case of uniform weights within each resultant.

The proposed intra-resultant weights given in Equation \eqref{eq:proposedweights} are not the only possible choices.  Weights could also be optimized for intermediate count rates.  In this case, slightly improved signal-to-noise ratios at higher count rates would come at the expense of slightly degraded signal-to-noise ratios at low count rates.  

Nonuniform weighting within resultants can improve the final signal-to-noise ratio at all count rates, significantly so for shorter exposures at low count rates.  The use of nonuniform weights will not affect the properties or removal of the correlated noise endemic to HxRG detectors, because these weights would still be the same for all pixels.  Cosmic ray flagging, discussed in \citetalias{Brandt_2024}, will similarly be unaffected in principle.  Sensitivitities to cosmic rays will change slightly, but a detailed analysis of that is beyond the scope of the current discussion.  A reweighting could complicate nonlinearity corrections though, if the nonlinear behavior is accurately known, this could be propagated through the known readout and weighting pattern.  

As shown in this section, nonuniform weighting within a resultant offers promise for preserving more useful information in a limited number of resultants.  In the read noise limit it can preserve all useful information.  If nonuniform weighting can be implemented in practice, it could improve the performance of a mission limited by downlink bandwidth.

\end{document}